\documentclass[aps,prd,superscriptaddress,nofootinbib]{revtex4}

\usepackage{scalerel}

\newcommand{\udt}[3]{#1^{#2}_{\phantom{#2}#3}}

\newcommand{\dut}[3]{#1_{#2}^{\phantom{#2}#3}}
\newcommand{\dudt}[4]{#1_{#2\phantom{#3}#4}^{\phantom{#2}#3}}
\newcommand{\udut}[4]{#1^{#2\phantom{#3}#4}_{\phantom{#2}#3}}

\usepackage{amsmath,amssymb,amsfonts,mathtools}
\usepackage{graphicx,epsfig}
\usepackage{hyperref}
\usepackage{bm,color}
\usepackage{accents}
\usepackage{tabularx,siunitx}
\usepackage{wasysym}

\newcommand{\lc}[1]{\accentset{\circ}{#1}}
\newcommand{\tg}[1]{\accentset{\land}{#1}}
\newcommand{\stg}[1]{\accentset{\scaleto{\diamond}{3.5pt}}{#1}}

\begin{document}

\title{The $3+1$ Formalism in the Geometric Trinity of Gravity}

\author{Salvatore Capozziello}
\email[]{capozziello@unina.it}
\affiliation{Dipartimento di Fisica  ``E. Pancini", Universit\`a di Napoli  ``Federico II", Via Cinthia, I-80126, Napoli, Italy.}
\affiliation{Istituto Nazionale di Fisica Nucleare (INFN), sez. di Napoli, Via Cinthia 9, I-80126 Napoli, Italy.}
\affiliation{Scuola Superiore Meridionale, Largo San Marcellino 10, 80138 Napoli, Italy.}
\affiliation{Laboratory for Theoretical Cosmology, Tomsk State University of Control Systems and Radioelectronics (TUSUR), 634050 Tomsk, Russia.}

\author{Andrew Finch}
\email[]{andrew.finch@um.edu.mt}
\affiliation{Institute of Space Sciences and Astronomy, University of Malta, Malta.}
\affiliation{Department of Physics, University of Malta, Malta.}

\author{Jackson Levi Said}
\email[]{jackson.said@um.edu.mt}
\affiliation{Institute of Space Sciences and Astronomy, University of Malta, Malta.}
\affiliation{Department of Physics, University of Malta, Malta.}

\author{Alessio Magro}
\email[]{alessio.magro@um.edu.mt}
\affiliation{Institute of Space Sciences and Astronomy, University of Malta, Malta.}
\affiliation{Department of Physics, University of Malta, Malta.}

\date{\today}

\begin{abstract}
The geometric trinity of gravity offers a platform in which gravity can be formulated in three analogous approaches, namely curvature, torsion and nonmetricity. In this vein, general relativity can be expressed in three dynamically equivalent ways which may offer insights into the different properties of these decompositions such as their Hamiltonian structure, the efficiency of numerical analyses, as well as the classification of gravitational field degrees of freedom. In this work, we take a $3+1$ decomposition of the teleparallel equivalent of general relativity and the symmetric teleparallel equivalent of general relativity which are both dynamically equivalent to curvature based general relativity. By splitting the spacetime metric and corresponding tetrad into their spatial and temporal parts as well as through finding the Gauss-like equations, it is possible to set up a general foundation for the different formulations of gravity. Based on these results, general $3$-tetrad and $3$-metric evolution equations are derived. Finally through the choice of the two respective connections, the metric $3+1$ formulation for general relativity is recovered  as well as the tetrad $3+1$ formulation of the teleparallel equivalent of general relativity and the metric $3+1$ formulation of symmetric teleparallel equivalent of general relativity. The approach is capable, in principle, of resolving common features of the various formulations of general relativity at a fundamental level and pointing out characteristics that extensions and alternatives to the various formulations can present.
\end{abstract}

\maketitle

\section{Introduction}

In light of the recent observational discovery of gravitational waves (GW), the need for  further work on  methods to improve numerical relativity simulations has become all the more pressing \cite{Abbott:2016blz}. Adding to this, the subsequent multimessenger discovery of a binary neutron star merger both in the GW spectrum with GW170817 \cite{TheLIGOScientific:2017qsa} as well as having an associated electromagnetic counterpart GRB170817A \cite{Goldstein:2017mmi} has drastically intensified the search for alternative and possible more efficient ways to numerically simulate gravitational theories and general relativity (GR) in particular. Another critical question to answer is whether these numerical approaches can tolerate the plethora of proposals for modified theories of gravity today on the market \cite{Capozziello:2011et,Nojiri:2017ncd,  Clifton:2011jh, CANTATA:2021ktz, Capozziello:2021bki}. While interesting approaches have been introduced in standard GR \cite{Baumgarte:2010ndz,Baumgarte:2021skc}, it is also interesting to consider possible $3+1$ formulations of GR using analogous constructions of gravity which are dynamically equivalent to GR.

The prospect of supporting a different geometric basis on which to build gravitational theories was worked on by Einstein himself \cite{2005physics...3046U} in his attempt to unify gravitation with electromagnetism. Ultimately, it turned out that this was not possible with the degrees of freedom (DoF) that he was using but efforts continued throughout the decades to use both teleparallel gravity (TG) \cite{RevModPhys.48.393,Hayashi:1967se,Hayashi:1979qx}, which relies on torsion, and symmetric teleparallel gravity (STG) \cite{Nester:1998mp,Dialektopoulos:2019mtr}, which is based on nonmetricity.
In its current state, TG is built by a shift in the connection from the Levi-Civita $\udt{\lc{\Gamma}}{\rho}{\mu\nu}$ (over-circles denote all quantities that use the Levi-Civita connection in their calculation) to the teleparallel connection $\udt{\tg{\Gamma}}{\rho}{\mu\nu}$ (over-hats denote all quantities that use the teleparallel connection in their calculation). This represents a transformation from curvature- to torsion-based geometry in our gravitational theories. On the other hand, STG is also formed by an exchange of connections where a disformation tensor $\udt{\stg{L}}{\alpha}{\mu\nu}$ (over-diamonds denote all quantities that assume nonmetricity in their calculation) can be used to replace the curvature associated with the Levi-Civita connection with nonmetricity.

Tetrads of TG  replace the metric tensor  as the fundamental dynamical objects of the theory. Specifically, the tetrad $\udt{e}{A}{\mu}$ and spin connection $\udt{\omega}{A}{B\mu}$ pair replaces the gravitational potentials related to the metric. Together,  tetrad and spin connection pair represents the fundamental DoFs of any ensuing teleparallel theory. The tetrad represents the gravitational part of this dynamics while the spin connection is a flat connection  describing the local Lorentz frame  \cite{Bahamonde:2021gfp}. This pair also appear in GR but are hidden into the internal structure of the theory. TG is different in that it separates these objects out instead of using the metric tensor as its fundamental variable. One reason for this is that, in GR, the spin connection is not flat and may contain gravitational contributions while in TG the spin connection is exclusively of inertial  nature \cite{PereiraBook}.

The shift from curvature- to torsion-based theories of gravity is intrinsically coupled with the change from tensorial measures of curvature to different tensors that are based on torsion. To this end, the central role that the Riemann tensor plays in gravity has to be changed with another object, the torsion tensor, on which gravitational theories can be built. This happens  because all measures of curvature identically vanish for tensors that are based on the teleparallel connection which is curvature-less. In this context, we also build scalar invariants which can be used to construct gravitational actions. One such scalar is the torsion scalar $\tg{T}$ which can be shown to be equal to the standard  Ricci scalar $\lc{R}$ up to a total divergence term (see Sec.~\ref{sec:TG_technical}). This is a crucial property since it means that the Hilbert-Einstein action can be analogously related to a torsional action producing a dynamically equivalent theory. This is called the teleparallel equivalent of general relativity (TEGR). One would expect this equivalent to be limited to the classical regime of the theory.

TEGR has a number of positive features that make it advantageous with respect to its curvature-based counterpart \cite{Bahamonde:2021gfp}. Firstly, the strong equivalence principle does not appear as a fundamental requirement of the theory which may be beneficial for some quantum gravity theories, despite the principle still being observationally satisfied \cite{Voisin:2020lqi}. Another crucial aspect of TEGR, related to quantum gravity,  is that the inertial nature of the spin connection makes it more akin to regular field theory making  approaches to quantum gravity more amenable to successful constructions of gravitational theories \cite{Aldrovandi:2005mz,Aldrovandi:2006cy}. This is related to the property that TEGR naturally produces a Gibbons-Hawking-York boundary term in its action giving it a well-defined Hamiltonian expression \cite{Oshita:2017nhn}. Altogether, these properties make TEGR more malleable towards a quantum setting and produce a likeness to the regular Yang-Mills theories of particle physics. 

On the other hand, STG continues to sustain the metric tensor as its fundamental dynamical object \cite{Nester:1998mp,Coley:2019zld,Adak:2004uh,Jarv:2018bgs}. However, here one normally sets both curvature and torsional contributions to gravity to zero by a Lagrange multiplier method \cite{BeltranJimenez:2017tkd,BeltranJimenez:2018vdo}. Moreover, STG also takes advantage of the Palatini approach in its field equations. As with TG, STG also depends on new tensorial quantities to be defined in order to construct gravitational theories. In particular, STG depends on the nonmetricity tensor, $\stg{Q}_{\lambda\mu\nu} := \stg{\nabla}_{\lambda}g_{\mu\nu}$, which then is the sole contributor to the disformation connection and the eventual field equations both for the metric and the connection (see Sec.~\ref{sec:STG_technical}). Another critical point is that as we can define a TEGR in TG, we can also define a symmetric teleparallel equivalent of general relativity (STEGR) in STG \cite{Adak:2005cd,Adak:2006rx} where a nonmetricity scalar $\stg{Q}$ turns out to be equal to the Levi-Civita connection based Ricci scalar up to a boundary term. This will also be dynamically equivalent to standard GR but may host different behavior  in its quantum regime. Hence, GR, TEGR and STEGR form a trinity approach to gravity \cite{Jimenez:2019woj} for the same dynamical equations.

The $3+1$ decomposition is determined by taking into account the temporal and spatial parts of the covariant field equations so that they can be probed in a numerical setting. This was first explored in standard GR through the work by Arnowitt, Deser, and Misner (ADM) \cite{1959PhRv..116.1322A} which produced two evolution equations and two constraint equations from which a plethora of physical setting can be probed in numerical relativity. However, this approach has not been extensively studied in the novel scenarios described in these other geometric settings. On the other hand, some interesting work has been advanced. For instance, in TG, Refs.\cite{Blixt:2018znp,Blixt:2019mkt,Blixt:2020ekl,Hohmann:2020muq} have explored important parts of the decomposition of TEGR. On the other end of the spectrum, Refs.\cite{Jimenez:2021hai,Ariwahjoedi:2020wmo} explored important properties in the formulation of the $3+1$ decomposition for STEGR.

Throughout this, work Greek symbols will be used for global indices and capital Latin symbols will be used to denote local coordinate indices, respectively. Global spatial indices will be denoted by $\{i,j,k\}$, and $\{\tilde{i},\tilde{j},\tilde{k}\}$ will denote local spatial indices. Finally any temporal index will be denoted by $0$ for the global case and by $\tilde{0}$ for the local case. The work is broken down as follows.  In Sec.~\ref{sec:TG_STG_intro}, we briefly introduce both TG and STG formalisms together with their respective TEGR and STEGR descriptions of GR. We then present a general approach to the $3+1$ decomposition in Sec.~\ref{Sec:General_for} where the decomposition is setup for the full trinity of gravity without assuming a particular connection for gravitation. In Sec.~\ref{sec:specifics}, we finally present the TG and STG $3+1$ decompositions for TEGR and STEGR respectively.  In Sec.~\ref{sec:conclu}, we discuss and summarize our main results.

\section{Teleparallel Theories of Gravity} \label{sec:TG_STG_intro}

The trinity of gravity \cite{Jimenez:2019woj} provides a mechanism by which we can form the same classical field equations that can be produced using either curvature-, torsion- or nonmetricity-based formulations of gravity. In this section, we introduce the basis on which these different formulations of gravity are built as well as laying out any nuances of the theories themselves.

\subsection{Teleparallel equivalent of general relativity}\label{sec:TG_technical}

TG represents a fundamental shift in the way that gravity is expressed in that the geometric curvature associated with the Levi-Civita connection $\lc{\Gamma}^{\sigma}_{\mu\nu}$ is now transformed into torsion associated with the corresponding teleparallel connection $\tg{\Gamma}^{\sigma}_{\mu\nu}$ \cite{Hayashi:1979qx}. The change in connection will alter the way that gravity is exhibited but not its geometric origins, meaning that we will continue to produce a metric tensor. However, the Riemann tensor, which produces a meaningful measure of curvature on a manifold (in standard gravity), will now identically vanish due to the curvature-less nature of the teleparallel connection, i.e. $\udt{\tg{R}}{\alpha}{\mu\nu\beta}\equiv 0$. A consequence of this result will be that a TG analog  has to be constructed in order to give a meaningful measure of torsion \cite{nakahara2003geometry,ortin2004gravity}.

In GR, the fundamental dynamical object is the metric tensor, $g_{\mu\nu}$, but in TG this becomes a derived quantity and its role is replaced by tetrad $\udt{e}{A}{\mu}$ and spin connection $\udt{\omega}{A}{B\mu}$ pairs. These act as a soldering agent between the general manifold (Greek indices) and the Minkowski space (Latin indices) \cite{Aldrovandi:2013wha}. In this context, the tetrads can be used to transform to (and from) the Minkowski metric through
\begin{align}\label{metric_tetrad_rel}
    \eta_{ab}=\dut{e}{A}{\mu}\dut{e}{B}{\nu}g_{\mu\nu}& &g_{\mu\nu}=\udt{e}{A}{\mu}\udt{e}{B}{\nu}\eta_{AB}\,.
\end{align}
The tetrads also observe the inverse conditions
\begin{align}
    \udt{e}{A}{\mu}\dut{e}{B}{\mu}=\delta^A_B& &\udt{e}{A}{\mu}\dut{e}{A}{\nu}=\delta^{\nu}_{\mu}\,,
\end{align}
for consistency.

The teleparallel connection is the most general linear affine connection that is both curvature-less and satisfies the metricity condition \cite{Aldrovandi:2013wha}, and can be defined as \cite{Weitzenbock1923}
\begin{equation}\label{tetrad_defs}
    \tg{\Gamma}^{\sigma}_{\nu\mu}:=\dut{e}{A}{\sigma}\partial_{\mu}\udt{e}{A}{\nu} + \dut{e}{A}{\sigma}\udt{\omega}{A}{B\mu}\udt{e}{B}{\nu}\,.
\end{equation}
The spin connection also appears in GR but it is mainly hidden within the internal structure of the theory \cite{misner1973gravitation}. Moreover, in TG, the spin connection is an inertial quantity and appears to preserve the covariance of the field equations \cite{Krssak:2015oua}, i.e. this acts to balance the potentially infinite solutions of Eq.(\ref{tetrad_defs}) with the necessity of covariance of the dynamical equations \cite{Krssak:2015oua}. In this way, the spin connection incorporates the DoFs associated with the local Lorentz transformation invariance of the theory \cite{Krssak:2018ywd}.

Considering the full breadth of Local Lorentz Transformations (LLTs), i.e.  Lorentz boosts and rotations, $\udt{\Lambda}{A}{B}$, the spin connection can be fully described by $\udt{\omega}{A}{B\mu}=\udt{\Lambda}{A}{C}\partial_{\mu}\dut{\Lambda}{B}{C}$ \cite{Aldrovandi:2013wha}. Thus, for the infinite number of solutions of Eq.(\ref{tetrad_defs}) (which arise due to the invariance of the metric under LLTs), the spin connection is produced by the field equations to sustain their general covariance under both diffeomorphisms and LLTs. It is for this reason that the spin connection components are representing the DoFs of the Lorentz group rather than extra DoFs of the theory \cite{Ferraro:2016wht,Bahamonde:2021gfp}. Thus, it is the combination of the tetrad choice and its associated spin connection that retains the covariance of TG. On the other hand, a frame can always be chosen in which the spin connection  vanishes since there will always exist a local frame that renders this result, which is called the Weitzenb"{o}ck gauge.  Given the teleparallel connection, the torsion tensor can be defined as \cite{Cai:2015emx}
\begin{equation}
    \udt{\tg{T}}{\sigma}{\mu\nu} := \tg{\Gamma}^{\sigma}_{[\nu\mu]}\,,
\end{equation}
where the square brackets denote the anti-symmetric operator. Here it should be noted that, throughout this paper, we define $A_{(\mu\nu)}=A_{\mu\nu}+A_{\nu\mu}$ and $A_{[\mu\nu]}=A_{\mu\nu}-A_{\nu\mu}$. From a dynamical point of view, the torsion tensor replaces the Riemann tensor: it gives a measure of torsion and thus geometric deformation in the theory. This also acts as the field strength of gravitation in TG, and transforms covariantly under both diffeomorphisms and LLTs. It is also convenient to define two other tensorial quantities that render a more concise form of the ensuing theory. First, consider the contorsion tensor which represents the difference between the teleparallel and Levi-Civita connections, namely
\begin{equation}
    \udt{\tg{K}}{\sigma}{\mu\nu} := \tg{\Gamma}^{\sigma}_{\mu\nu} - \lc{\Gamma}^{\sigma}_{\mu\nu} =\frac{1}{2}\left(\dudt{\tg{T}}{\mu}{\sigma}{\nu} + \dudt{\tg{T}}{\nu}{\sigma}{\mu} - \udt{\tg{T}}{\sigma}{\mu\nu}\right)\,.
\end{equation}
This tensor plays a crucial role in relating results in TG with their analogue in standard gravity. The other ingredient in TG is the so-called superpotential
\begin{equation}
    \dut{\tg{S}}{A}{\mu\nu}:=\udt{\tg{K}}{\mu\nu}{A} - \dut{e}{A}{\nu}\udt{\tg{T}}{\alpha\mu}{\alpha} + \dut{e}{A}{\mu}\udt{\tg{T}}{\alpha\nu}{\alpha}\,,
\end{equation}
which is linked with the gauge current representation of the gravitational energy-momentum tensor within TG \cite{Aldrovandi:2003pa}, but this remains an open issue in TG \cite{Koivisto:2019jra,Capozziello:2018qcp}.

Contracting the torsion tensor with its superpotential produces the torsion scalar
\begin{equation}
    \tg{T}:=\frac{1}{2}\dut{\tg{S}}{A}{\mu\nu}\udt{\tg{T}}{A}{\mu\nu}\,,
\end{equation}
which is calculated entirely on the teleparallel connection in an analogous way as the Ricci scalar depends only on the Levi-Civita connection. By choosing to construct the torsion scalar in this way, it turns out that the torsion and Ricci scalars are equivalent up to a boundary term \cite{Bahamonde:2015zma,Farrugia:2020fcu,Capozziello:2018qcp}
\begin{equation}
    \tg{R}=\lc{R} + \tg{T} -\frac{2}{e}\partial_{\mu}\left(e\udut{\tg{T}}{\sigma}{\sigma}{\mu}\right) = 0\,,
\end{equation}
where $R$ is the Ricci scalar in terms of the teleparallel connection, and $\mathring{R}$ is the regular Ricci scalar from standard gravity (calculated by the Levi-Civita connection). The first conclusion of this observation is that the Lagrangian of the Hilbert-Einstein action will be equal to the torsion scalar up to a boundary term \cite{Aldrovandi:2013wha,Capozziello:2018qcp}
\begin{equation}
    \lc{R} = -\tg{T} + \frac{2}{e}\partial_{\mu}\left(e\udut{\tg{T}}{\sigma}{\sigma}{\mu}\right) := -\tg{T} + \tg{B}\,,
\end{equation}
where $e=\det\left(\udt{e}{a}{\mu}\right)=\sqrt{-g}$. This equivalence alone will insure that both scalars will produce the same dynamical equations of motion. In this way, the TEGR action can be written as
\begin{equation}\label{TEGR_action}
    \mathcal{S}_{\text{TEGR}} = -\frac{1}{2\kappa^2}\int d^4 x\; e\tg{T} + \int d^4 x\; e\mathcal{L}_{\text{m}}\,,
\end{equation}
where $\kappa^2=8\pi G$ and $\mathcal{L}_{\text{m}}$ is the matter Lagrangian. Both TEGR and the Hilbert-Einstein actions lead to the same dynamical equations, but they differ in their Lagrangians by a boundary term which, in GR, produces fourth-order terms in extensions beyond GR, such as in $f(\lc{R})$ gravity \cite{Capozziello:2002rd,Sotiriou:2008rp,Faraoni:2008mf,Capozziello:2011et}. Thus, in TG, the second- and fourth-order contributions from the action are somewhat decoupled in the torsion scalar and boundary term. This is not relevant for comparisons of GR and TEGR at the level of their field equations, but  becomes an active agent when modifications are considered \cite{Cai:2015emx}.

In GR, the Hilbert-Einstein action leads to Einstein's field equations (EFFs) where $\lc{G}_{\mu\nu} = \kappa^2 \Theta_{\mu\nu}$, and where the Einstein tensor is determined by the Levi-Civita connection \cite{misner1973gravitation}, and where the regular energy-momentum tensor is defined as $\Theta_{\mu\nu} := -2/\sqrt{-g}\, \delta \left(\sqrt{-g}\mathcal{L}_{\rm m}\right)/\delta g^{\mu\nu}$. The TEGR manifestation of TG leads to the TEGR action in Eq.(\ref{TEGR_action}), where a variation with respect to the tetrad leads directly to \cite{Krssak:2015oua}
\begin{align}\label{TEGR_FEs}
    \lc{G}_{\mu\nu} \equiv \tg{G}_{\mu\nu} &:= e^{-1}e^{A}{}_{\mu}g_{\nu\rho}\partial_\sigma(e \tg{S}_a{}^{\rho\sigma})-\tg{S}_{B}{}^{\sigma}{}_{\nu}\tg{T}^{B}{}_{\sigma\mu}\\
    & +\frac{1}{4}\tg{T} g_{\mu\nu}-e^{A}{}_\mu \omega ^{B}{}_{A\sigma}\tg{S}_{B\nu}{}^{\sigma} = \kappa^2 \Theta_{\mu\nu}\,\nonumber,
\end{align}
which are dynamically equivalent to the EFFs except that they are based on the tetrad rather than on the metric tensor. Moreover, the TEGR field equations do not contain some of the non-dynamical boundary term contributions of the Ricci scalar \cite{Cai:2015emx}. In all other TG theories of gravity, a variation with respect to the spin connection leads to a separate set of field equations representing the six DoFs associated with the Lorentz group. However, in the very special case of TEGR, these equations are identically satisfied and so do not play an active role in the dynamics of the theory \cite{Bahamonde:2021gfp}.

\subsection{Symmetric teleparallel equivalent of general relativity}\label{sec:STG_technical}

Analogously, STG embodies the transition from curvature-based geometric gravity to one based on the property of nonmetricity. In this context, the curvature and torsion associated with the theory are set to vanish so that only nonmetricity embodies the effects of gravity. Thus, we can define a nonmetricity tensor through
\begin{equation}
    \stg{Q}_{\lambda\mu\nu} := \stg{\nabla}_\lambda g_{\mu\nu}\,,
\end{equation}
which uses the inverse metric by appropriate contractions to give
\begin{equation}
    \dut{\stg{Q}}{\lambda}{\mu\nu} = -\stg{\nabla}_\lambda g^{\mu\nu}\,,
\end{equation}
and where $\stg{Q}^{\alpha\mu\nu}=-g^{\alpha\beta}\stg{\nabla}_{\beta}g^{\mu\nu}$.
Different to TG, STG retains the metric as the fundamental dynamical object of the theory meaning that the transition is not so dramatic. 

Another way to view the trinity of gravity is through the general linear affine connection given by \cite{ortin2004gravity,Hehl:1994ue}
\begin{equation}
    \udt{\Gamma}{\alpha}{\mu\nu} = \udt{\lc{\Gamma}}{\alpha}{\mu\nu} + \udt{\tg{K}}{\alpha}{\mu\nu} + \udt{\stg{L}}{\alpha}{\mu\nu}\,,\label{eq:gen_connec}
\end{equation}
where $\udt{\stg{L}}{\alpha}{\mu\nu}$ represents the disformation tensor which encodes the contribution of the nonmetricity tensor \cite{Nester:1998mp,BeltranJimenez:2017tkd}, and is defined as
\begin{equation}
    \udt{\stg{L}}{\alpha}{\mu\nu} := \frac{1}{2}g^{\lambda\alpha}\left(Q_{\lambda\mu\nu}-Q_{\mu\lambda\nu}-Q_{\nu\lambda\mu}\right)\,,\label{eq:disf}
\end{equation}
which shares a number of symmetries with the Levi-Civita connection. The class of STG is given by the conditions of vanishing curvature
\begin{equation}
    \udt{\stg{R}}{\rho}{\sigma\mu\nu}\equiv0\,,
\end{equation}
and vanishing torsion
\begin{equation}
    \udt{\stg{T}}{\rho}{\mu\nu}\equiv0\,,
\end{equation}
which represent curvature- and torsion-based constructions of gravity, respectively. The most general connection that satisfies  conditions of this kind is the symmetric teleparallel connection
\begin{equation}
    \stg{\Gamma}^{\alpha}{}_{\mu\nu}:=\frac{\partial  x^{\alpha}}{\partial\xi^{\sigma}}\frac{\partial^{2}\xi^{\sigma}}{\partial x^{\mu}\partial  x^{\nu}}\,,\label{eq:stg_conn}
\end{equation}
where $\xi^{\sigma}=\xi^{\sigma}(x)$ is an arbitrary function of spacetime position. It turns out that this connection can be derived from vanishing connection components through the coordinate transformation
\begin{equation}
    x^{\mu}\rightarrow\xi^{\mu}(x^{\nu})\,.
\end{equation}
In this setting, the connection \eqref{eq:stg_conn} turns out to be a pure gauge connection, and hence it is always possible to determine a coordinate transformation in which this vanishes. This is called the coincident gauge \cite{BeltranJimenez:2017tkd}, and can be useful in simplifying calculations.

Thus, it is possible to construct a gravitational theory based solely on the disformation tensor where the effect of gravity is communicated through measures of nonmetricity rather than curvature or torsion. This is analogous to the way that the regular Riemann tensor is based only on the Levi-Civita connection. One of these theories is STEGR which is expressed through the Lagrangian
\begin{equation}
    \mathcal{\stg{L}}_{\text{STEGR}}=\frac{\sqrt{-g}}{16\pi G} \stg{Q}\,,\label{lag_stg}
\end{equation}
where 
\begin{equation}
    \stg{Q} = g^{\mu\nu}\left(\udt{\stg{L}}{\alpha}{\beta\mu}\udt{\stg{L}}{\beta}{\nu\alpha}-\udt{\stg{L}}{ \alpha}{\beta\alpha}\udt{\stg{L}}{\beta}{\mu\nu}\right)\,,
\end{equation}
which is the nonmetricity scalar.

To better see the equivalence with GR, we can rewrite the Hilbert-Einstein Lagrangian in terms of the Levi-Civita connection giving \cite{Harko:2018gxr} 
\begin{equation}
    \mathcal{\lc{L}}_{HE} = \frac{\sqrt{-g}}{16\pi G}\lc{R} = \mathcal{\lc{L}}_{E} + \lc{B}\,,\label{EH}
\end{equation}
where $\mathcal{L}_{E}$ represents the Einstein Lagrangian contribution constructed by the Levi-Civita connection \cite{Einstein:1916cd} in which
\begin{equation}
    \mathcal{\lc{L}}_{E} := \frac{\sqrt{-g}}{16\pi G}\,g^{\mu\nu}\left(\udt{\lc{\Gamma}}{\alpha}{\beta\mu} \udt{\lc{\Gamma}}{\beta}{\nu\alpha} - \udt{\lc{\Gamma}}{\alpha}{\beta\alpha} \udt{\lc{\Gamma}}{\beta}{\mu\nu}\right)\,,\label{eq:ELag}
\end{equation}
and where the boundary term is defined by
\begin{equation}
    \lc{B} = \frac{\sqrt{-g}}{16\pi G}\,\left(g^{\alpha\mu}\lc{\nabla}_{\alpha}\lc{\Gamma}^{\nu}{}_{\mu\nu}-g^{\mu\nu}\lc{\nabla}_{\alpha}\lc{\Gamma}^{\alpha}{}_{\mu\nu}\right)\,,\label{boundary_ein_def}
\end{equation}
which is a total divergence term. The Hilbert-Einstein Lagrangian completes the original Einstein Lagrangian in that it adds the boundary term $\lc{B}$ which renders the theory covariant \cite{ortin2004gravity}.

Now, assuming the coincident, where the connection vanishes ($\udt{\stg{\Gamma}}{\alpha}{\mu\nu} \equiv 0$), means that the covariant derivative directly reduces to the ordinary partial derivative ($\stg{\nabla}_{\mu} \rightarrow \partial_{\mu}$), it follows that the disformation tensor turns out to be the negative of the Christoffel symbols
\begin{equation}
    \udt{\stg{L}}{\alpha}{\mu\nu} = -\udt{\lc{\Gamma}}{\alpha}{\mu\nu}\,.
\end{equation}
Thus, it follows that the field equations of the STEGR Lagrangian must produce the exact same field equations as GR despite being based on nonmetricity as the means by which geometric deformation takes place \cite{Conroy:2017yln}. Hence, both GR and STEGR turn out to be dynamically equivalent, as do TEGR and STEGR.

The Einstein Lagrangian \eqref{eq:ELag} necessitates a boundary term in order to be diffeomorphism invariant, while in STEGR any arbitrary coordinate transformation of the Lagrangian \eqref{lag_stg} remains invariant. Thus STEGR retains diffeomorphism invariance without the boundary term. This is true even if the coincident gauge is not considered \cite{Koivisto:2019jra,Koivisto:2018aip}. In the same way that the Hilbert-Einstein action is a covariantization of the Einstein action, the STEGR action is another way to covariantize this action. While the Hilbert-Einstein action achieves this by simply adding a boundary term, STEGR makes the nontrivial shift to the symmetric teleparallel connection to accomplish this result.

The STEGR action thus turns out to be given by \cite{Victor_stg, BeltranJimenez:2018vdo}
\begin{align}
    S_{G} &= \int d^{4}x\,\Big[\frac{\sqrt{-g}}{2\kappa^2}\,\stg{Q} + \sqrt{-g}\mathcal{L}_{m}\Big]\,,\label{eq:stegr_action}
\end{align}
which naturally leads to the conjugate to the STEGR Lagrangian through
\begin{align}
     \udt{\stg{P}}{\alpha}{\mu\nu} &:= \frac{1}{2\sqrt{-g}} \frac{\partial\left(\sqrt{-g} \stg{Q}\right)}{\partial \dut{\stg{Q}}{\alpha}{\mu\nu}}\nonumber\\
    &= \frac{1}{4}\udt{\stg{Q}}{\alpha}{\mu\nu} - \frac{1}{4}\dudt{\stg{Q}}{(\mu}{\alpha}{\nu)} - \frac{1}{4} g_{\mu\nu}\udt{\stg{Q}}{\alpha\beta}{\beta} + \frac{1}{4}\left[\dut{\stg{Q}}{\beta}{\beta\alpha} g_{\mu\nu} + \frac{1}{2}\udt{\delta}{\alpha}{(\mu}\dudt{\stg{Q}}{\nu)}{\beta}{\beta} \right]\,,
\end{align}
which provides a convenient alternative to describing the metricity scalar as $\stg{Q} = - \stg{Q}_{\alpha\mu\nu} \stg{P}^{\alpha\mu\nu}$ \cite{Soudi:2018dhv}. By taking a variation of the action in Eq.~\eqref{eq:stegr_action} with respect to the metric tensor, the field equations are derived  \cite{Victor_stg}
\begin{equation}\label{eq:metric_FEs}
    2\stg{\nabla}_{\alpha}\left(\sqrt{-g} \udt{\stg{P}}{\alpha}{\mu\nu}\right) - q_{\mu\nu} - \frac{\sqrt{-g}\,\stg{Q}}{2}\,g_{\mu\nu} = \kappa^2\sqrt{-g}\Theta_{\mu\nu}\,,
\end{equation}
where
\begin{equation}
    \frac{1}{\sqrt{-g}} q_{\mu\nu} = \frac{1}{4}\left(2\stg{Q}_{\alpha\beta\mu}\udt{\stg{Q}}{\alpha\beta}{\nu} - \stg{Q}_{\mu\alpha\beta}\dut{\stg{Q}}{\nu}{\alpha\beta} \right) - \frac{1}{2} \stg{Q}_{\alpha\beta\mu}\udt{\stg{Q}}{\beta\alpha}{\nu} - \frac{1}{4}\left(2\dudt{\stg{Q}}{\alpha}{\beta}{\beta}\udt{\stg{Q}}{\alpha}{\mu\nu} - \dudt{\stg{Q}}{\mu}{\beta}{\beta}\dudt{\stg{Q}}{\nu}{\beta}{\beta}\right) + \frac{1}{2} \dudt{\stg{Q}}{\beta}{\beta}{\alpha}\udt{\stg{Q}}{\alpha}{\mu\nu}\,.
\end{equation}
The connection, being independent of the metric in this scenario, also produces independent field equations. By assuming vanishing hypermomentum \cite{BeltranJimenez:2018vdo}, the variation with respect to the connection produces the connection field equation
\begin{equation}\label{eq:connection_FEs}
    \stg{\nabla}_{\mu}\stg{\nabla}_{\nu} \left(\sqrt{-g} \udt{\stg{P}}{\mu\nu}{\alpha}\right) = 0\,.
\end{equation}
Together, the metric field Eqs.\eqref{eq:metric_FEs} and connection field Eqs.\eqref{eq:connection_FEs} represent the total DoFs of the dynamics in STEGR. Moreover, for STEGR, the connection equation is trivially satisfied when the coincident gauge is utilized.

\section{The 3+1 formulation for a general linear affine Connection without Metricity}\label{Sec:General_for}

Let us now formulate two new approaches to the $3+1$ decomposition of gravity, one tetrad-based and the other metric-based. This is achieved by considering a general affine connection, $\Gamma^{\lambda}_{\mu\nu}$, without metricity (tensors, and covariant derivatives with no accents denote all quantities that use this general connection in their calculation). This allows the general forms of curvature, torsion, spin and nonmetricity terms to feature in the ensuing equations. Throughout this section, we show that the metric and tetrad formalisms are each other consistent  at every level. It should be noted that, from this point onwards, $3$-tensors that have the same symbol as their 4 dimensional counterparts will be denoted with a $(3)$ superscript or subscript, and that purely local tensors will be denoted with a `$\;\tilde{\;}\;$' annotation to keep consistent with the index notation defined above.

\subsection{Basic definitions}

To set up a basis for the decomposition strategy being developed in this section, we present some definitions and some preliminary relationships between the main variables of the formalism. To this end, we let $M$ be a $4$-dimensional manifold with a global spacetime metric $g_{\mu\nu}$ which is derived from a tetrad $e^A_{\;\;\nu}$ such that 
\begin{equation}\label{eq.local_global_tetrad}
    g_{\mu\nu}=\eta_{AB}e^A_{\;\;\mu} e^B_{\;\;\nu}.
\end{equation}
We assume that ($M$, $g_{\mu\nu}$) can be foliated into non-intersecting spacelike $3$-surfaces, $\Sigma$, such that each surface is a level surface of a scalar function "$t$". This implies that these slices are purely spatial slices each at their own instance in time. In this case, this function can be interpreted as a global time function.

The one-form $\Omega_\nu$ is now defined as the covariant derivative of this time function, $\nabla_\nu{t}$. By contracting this one-form twice with the metric of the manifold,  we obtain the norm of $\Omega$ \cite{baumgarte},
\begin{align}
    {|\Omega|}^2&=g^{\mu\nu}\nabla_\mu{t}\nabla_\nu{t}\nonumber\\
    &=g^{00}\nonumber\\
    &=-\frac{1}{\alpha^2}\,,
\end{align}
where $\alpha$ is the lapse function and $\nabla_\mu$ is the general covariant derivative. The lapse function can be interpreted as a measure of how much time elapses between one time slice and the next. It is also assumed that $\alpha$ is positive making $\Omega_\nu$ timelike and the hypersurfaces, $\Sigma$, everywhere spacelike. 

The used general covariant derivative  is associated with the general affine connection. This derivative also adheres to nonmetricity \cite{Teleparallel_Palatini,Victor_stg}
\begin{align}
    \nabla_\lambda g_{\mu\nu} =& Q_{\lambda\mu\nu}\,,\\
    \nabla_\lambda g^{\mu\nu} =& -{Q_{\lambda}}^{\mu\nu}\,.
\end{align}
In general, the following convention for the covariant derivative of any mixed tensor $A^{a\nu}_{\;\;\;\;b\mu}$ will be followed throughout this work
\begin{align}
\nabla_\lambda A^{A\nu}_{\;\;\;\;b\mu}=A^{C\rho}_{\;\;\;\;B\mu}&\omega^A_{\;\;C\lambda}+A^{C\rho}_{\;\;\;\;B\mu}\Gamma^\nu_{\rho\lambda}\\\nonumber
&-A^{A\rho}_{\;\;\;\;C\mu}\omega^C_{\;\;B\lambda}-A^{A\nu}_{\;\;\;\;B\rho}\Gamma^\rho_{\mu\lambda}\,.
\end{align}
\noindent Here $\omega^A_{\;\;C\lambda}$ is the local spin connection and $\Gamma^\nu_{\rho\lambda}$ is a global general affine connection. By normalizing $\Omega$, the unit normal to the foliations and its inverse are defined as
\begin{align}
    n_\nu&:=-\alpha\Omega_\nu,\label{eq:ndef}\\
    n^\nu&:=-g^{\mu\nu}(\alpha\Omega_\mu)\label{eq:invndef}\\
    &=-g^{\mu\nu}(\alpha\nabla_\mu{\left(t\right)})\,,\nonumber
\end{align}
such that 
\begin{equation}\label{eq:ortogofn}
    n^{\nu}n_\nu=-1\,.
\end{equation}
The normal tensors are built to generate this negative sign so as to point the normal $n^\nu$ towards increasing time $t$. 

We can now set our normal vector and co-vector in terms of the lapse function $\alpha$ and the shift vector $\beta$ as in standard gravity as \cite{baumgarte,gourgoulhon},
\begin{align}
    n_\mu=&\left( -\alpha,0,0,0\right)\,,\\
    n^\mu=&\left(\frac{1}{\alpha},-\frac{1}{\alpha}\beta^i\right)\,.
\end{align}
Having thoroughly defined the normal to the foliations, the $3$-metric on the foliation $\Sigma$ can now be set to
\begin{align}
    \gamma^{\mu\nu}=&g^{\mu\nu}+n^{\mu}n^{\nu}\,,\label{eq. metric Split}\\
    \gamma_{\mu\nu}=&g_{\mu\nu}+n_{\mu}n_{\nu}\,.\label{eq. inv metric Split}
\end{align}
In order to break up arbitrary 4-dimensional tensors on the manifold ($M$,$g_{\mu\nu}$) into their spatial and temporal parts, a variant decomposition of this metric, $g_{\mu\nu}\gamma^{\lambda\mu}$, can be used. This is defined as follows 
\begin{equation}\label{eq. 3-delta Split}
    \gamma^\mu_{\;\;\nu}=\delta^\mu_\nu+n^{\mu}n_{\nu}\,.
\end{equation}
With this definition in hand, $\gamma^\mu_{\;\;\nu}$ can be thought  as the $3$-delta. While this form of the delta function exhibits all the properties of a spatial delta on spatial vectors and tensors, this is not the spatial identity matrix. This tensor has off-diagonal temporal components similarly to the inverse spatial metric.
\begin{align}
    \gamma^{\mu\nu}=&\begin{pmatrix}
    0 & 0\\
    0 & \gamma^{ij}
    \end{pmatrix}\,,\\
    \gamma_{\mu\nu}=&\begin{pmatrix}
    \beta^k \beta_k & \beta_i\\
    \beta_j & \gamma_{ij} 
    \end{pmatrix}\,,\\
    {\gamma^\mu}_\nu=&\begin{pmatrix}
    0 & \beta^i\\
    0 & {\gamma^i}_j 
    \end{pmatrix}\,.
\end{align}
As a result of this and Eqs.~(\ref{eq. metric Split}, \ref{eq. inv metric Split}, \ref{eq. 3-delta Split}), we get 
\begin{align}
    g^{\mu\nu}=&\begin{pmatrix}
    -\frac{1}{\alpha^2} & \frac{1}{\alpha^2}\beta^i\\
    \frac{1}{\alpha^2}\beta^i & \gamma^{ij}- \frac{1}{\alpha^2}\beta^i\beta^j
    \end{pmatrix}\,,\label{eq. metric components}\\
    g_{\mu\nu}=&\begin{pmatrix}
    -\alpha^2 + \beta^k \beta_k & \beta_i\\
    \beta_j & \gamma_{ij} 
    \end{pmatrix}\,.\label{eq. inv metric components}
\end{align}
The following relations are some useful consequences of the previous definitions
\begin{align}\label{eq:nV0}
    \gamma^{\mu\lambda}\gamma_{\lambda\nu}&={\gamma^{\mu}}_\nu\,,\\
    \gamma^\mu_{\;\;\nu}V^{(3)}_{\mu}&=V^{(3)}_{\nu}\,,\\
    n^{\mu}V^{(3)}_{\mu}&=0\,,\\
    n^{\mu}\gamma_{\mu\nu}&=0\,.
\end{align}

Similar to the the global manifold defined above, we now specify an inertial manifold $\tilde{M}$ along with its metric, the Minkowski metric $\eta_{ab}$, and non-intersecting spacelike $3$-surfaces $\tilde{\Sigma}$ are defined. The normal $\tilde{n}^a$ to the foliations and the inertial $3$-metric, $\tilde{\gamma}_{ab}$, can also be similarly defined.

It should be noted that since we are assuming nonmetricity, the covariant derivative of the Minkowski metric is non-zero in this setting, and it results in
\begin{align}
    \nabla_\lambda \eta_{AB}& = \partial_\lambda \eta_{AB} - \eta_{CB}{\omega^{C}}_{A\lambda}- \eta_{AC}{\omega^{C}}_{B\lambda}\\
    &= - \eta_{CB}{\omega^{C}}_{A\lambda}- \eta_{AC}{\omega^{C}}_{B\lambda}\,.\nonumber
\end{align}
In theories with metricity, this would result in an antisymmetry in the first two indices of the spin connection.

By choosing the Minkowski metric as our local metric, we can show that the inertial normal $\tilde{n}$ and its inverse are constant vectors
\begin{align}
    \tilde{n}_A&=\begin{pmatrix}-1&0&0&0\end{pmatrix}\,,\label{eq:inerital-nelem}\\
    \tilde{n}^A&=\begin{pmatrix}1&0&0&0\end{pmatrix}\,,\label{eq:inerital-invnelem}
\end{align}
since the lapse function and the shift vector are unity and a zero vector respectively, in this case. If a non-diagonal local metric were chosen, the elements of $\tilde{n}^a$ would become dependent on the spatial shift vector $\tilde{\beta}^{\tilde{i}}$, the same as the standard global one  \cite{baumgarte}. 

At this point we reintroduce the tetrad and attempt to find a form for it equivalent to the metric ones in Eqs.~(\ref{eq. metric components},\ref{eq. inv metric components}). Starting from Eq.~(\ref{eq.local_global_tetrad}) we get that 
\begin{equation}
    g_{\mu\nu}=\begin{pmatrix}
    {e^{\tilde{0}}}_{0}{e^{\tilde{0}}}_{0}\eta_{\tilde{0}\tilde{0}}+{e^{\tilde{i}}}_{0}{e^{\tilde{j}}}_{0}\eta_{\tilde{i}\tilde{j}}  &{e^{\tilde{0}}}_{i}{e^{\tilde{0}}}_{0}\eta_{\tilde{0}\tilde{0}}+{e^{\tilde{i}}}_{i}{e^{\tilde{j}}}_{0}\eta_{\tilde{i}\tilde{j}}\\
    {e^{\tilde{0}}}_{0}{e^{\tilde{0}}}_{j}\eta_{\tilde{0}\tilde{0}}+{e^{\tilde{i}}}_{0}{e^{\tilde{j}}}_{j}\eta_{\tilde{i}\tilde{j}}  &{e^{\tilde{0}}}_{i}{e^{\tilde{0}}}_{j}\eta_{\tilde{0}\tilde{0}}+{e^{\tilde{i}}}_{i}{e^{\tilde{j}}}_{j}\eta_{\tilde{i}\tilde{j}}
    \end{pmatrix}\,.\label{eq:metric_decom}
\end{equation}
Comparing this to Eq.(\ref{eq. inv metric components}), we get that
\begin{equation}
    {e^{\tilde{0}}}_{i}{e^{\tilde{0}}}_{j}\eta_{\tilde{0}\tilde{0}}+{e^{\tilde{i}}}_{i}{e^{\tilde{j}}}_{j}\eta_{\tilde{i}\tilde{j}}=\gamma_{ij}\,.
\end{equation}
Now since we know that ${e^{\tilde{i}}}_{i}{e^{\tilde{j}}}_{j}\eta_{\tilde{i}\tilde{j}}=\gamma_{ij}$ and that $\eta_{\tilde{0}\tilde{0}}=-1$, we conclude that ${e^{\tilde{0}}}_{i}=0$. 

Using this we also get that $\beta_j={e^{\tilde{i}}}_{i}{e^{\tilde{j}}}_{0}\eta_{\tilde{i}\tilde{j}}$, from which we can conclude that ${e^{\tilde{i}}}_0={e^{\tilde{i}}}_m\gamma^{m n}\beta_{n}$. Substituting this in the second term of $g_{00}$ component of Eq.~\eqref{eq:metric_decom}, we get that ${e^{\tilde{i}}}_{0}{e^{\tilde{j}}}_{0}\eta_{\tilde{i}\tilde{j}}=\beta^k\beta_k$ implying that ${e^{\tilde{0}}}_{0}=\pm\alpha$. Noticing that $e_A^{\;\;\nu}=\eta_{AB} g^{\mu\nu} e^B_{\;\;\mu}$, while using the above result, we get
\begin{align}
    {e^A}_{\nu}=&\begin{pmatrix}
    \pm\alpha & {e^{\tilde{i}}}_k\beta^{k}\\
    0 & {e^{\tilde{i}}}_{j}\end{pmatrix},\\
    {e_A}^{\nu}=&\begin{pmatrix}
    \pm\frac{1}{\alpha} & 0\\
    \mp\frac{1}{\alpha}\beta^j & {e_{\tilde{i}}}^{j}\end{pmatrix}\,.
\end{align}
At this point, it is worth noticing that the two defined normal vectors, the local one and the global one, are not necessarily related through the tetrad, namely
\begin{equation}
    \tilde{n}_A = \dut{e}{A}{\nu} n_\nu\,,
\end{equation}
since this is only the case if ${e_{\tilde{i}}}^0=0$. From the above considerations, we  conclude that, for the current choice of metric Eq.~(\ref{eq. metric components}), this is  the case and so, using this relation and its variant decompositions, we obtain the correct signs for the tetrad components resulting in
\begin{align}
    {e^A}_{\nu}=&\begin{pmatrix}
    \alpha & {e^{\tilde{i}}}_k\beta^{k}\\
    0 & {e^{\tilde{i}}}_{j}\end{pmatrix}\,,\label{eq. tetrad components}\\
    {e_A}^{\nu}=&\begin{pmatrix}
    \frac{1}{\alpha} & 0\\
    -\frac{1}{\alpha}\beta^j & {e_{\tilde{i}}}^{j}\end{pmatrix}\,.\label{eq. inv tetrad components}
\end{align}
It has to be noted that these agree with the results obtained in Refs.~\cite{Blixt:2019mkt,Blixt:2020ekl} at least up to the local index $3+1$ split. Having the tatrads in this form, we can now move on to find an equivalent to $\gamma_{\mu\nu}$, say ${\theta^A}_\mu$, such that 
\begin{equation}
    {\theta^A}_\mu={e^A}_\mu+{U^A}_\mu\,,
\end{equation}
where ${U^A}_\mu$ is some tensor that will embody the temporal part of the tetrad. The only property we need for ${\theta^A}_\mu$ is that it is orthogonal to the normal vectors. Since we have two unknowns, ${\theta^A}_\mu$ and ${U^A}_\mu$, there is nothing preventing us from defining ${U^A}_\mu$ as any tensor we want so long as it preserves the orthogonality of ${\theta^A}_\mu$ with the normal vectors. Thus, we set this statement as
\begin{equation}
    {U^A}_\mu=\tilde{n}^A n_\mu\,.
\end{equation}
Below, we show that this property  preserves indeed the orthogonality of ${\theta^A}_\mu$ in that we get
\begin{align}
    n^\mu {\theta^A}_\mu =& n^\mu {e^A}_\mu +n^\mu n_\mu \tilde{n}^A \\
    =&\tilde{n}^A - \tilde{n}^A\nonumber\\
    =&0\,,\nonumber\\
    n_A {\theta^A}_\mu =& \tilde{n}_A {e^A}_\mu +n^\mu \tilde{n}_A \tilde{n}^A \\
    =&n^\mu - n^\mu\nonumber\\
    =&0\,.\nonumber
\end{align}
The same exact procedure can be applied when splitting the inverse tetrad, giving
\begin{align}
    {\theta^A}_\mu={e^A}_\mu+{\tilde{n}^A}n_\mu\,,\\
    {\theta_A}^\mu={e_A}^\mu+{\tilde{n}_A}n^\mu\,.
\end{align}
Using these equations, we can now determine the components of the  $\theta$ tensors to be
\begin{align}
    {\theta^A}_{\nu}=&\begin{pmatrix}
    0 & {e^{\tilde{i}}}_k\beta^{k}\\
    0 & {e^{\tilde{i}}}_{j}\end{pmatrix}\,,\label{eq. theta components}\\
    {\theta_A}^{\nu}=&\begin{pmatrix}
    0 & 0\\
    0 & {e_{\tilde{i}}}^{j}\end{pmatrix}\,.\label{eq. inv theta components}
\end{align}
Among others, some convenient properties of this form of ${\theta^A}_\mu$ that one can  easily show are
\begin{align}
    {\theta^A}_\mu{\theta_A}^\nu&= {\gamma^\nu}_\mu\,,\\
    {\theta^A}_\mu{\theta_B}^\mu&={\tilde{\gamma}^A}_B\,,\\
    \tilde{\gamma}_{AB}\theta^A_{\;\;\mu} \theta^B_{\;\;\nu}&=\gamma_{\mu\nu}\,,\\
    \tilde{\gamma}^{AB}\theta_A^{\;\;\mu} \theta_B^{\;\;\nu}&=\gamma^{\mu\nu}\,.
\end{align}
Through these properties,  we can then derive a number of other relations as a consequence. 

While $\theta^A_{\;\;\mu}$ should not be confused with ${e^{\tilde{i}}}_{j}$, it has to be noted that from this point onward we will be re-labeling ${\theta^{A}}_{\mu}$ as ${e^{A}}^{(3)}_{\mu}$ since it effectively embodies the role of the \textit{spatial tetrad}.

Let us  now define the \textit{spatial covariant derivative} of a general spatial tensor $A$ as
\begin{equation}
    D_{\nu}A^A_{\;\;\mu}=\gamma^\lambda_{\;\;\mu}\tilde{\gamma}^A_{\;\;B}\gamma^\rho_{\;\;\nu}\nabla_{\rho}A^B_{\;\;\lambda}\,.
\end{equation}
This derivative can be shown to follow the Leibniz rule only when applied to spatial tensors thus resulting in additional terms if applied to a spacetime tensor \cite{gourgoulhon}. Two results of note from this definition are the following
\begin{align}
    D_\nu n^\mu&=\partial^{(3)}_\nu n^\mu + n^\lambda \Gamma^{\mu(3)}_{\lambda \nu}\\
    &=\partial^{(3)}_\nu n^\mu\,,\nonumber\\
    D_\nu \tilde{n}^A &=\partial^{(3)}_\nu \tilde{n}^A + \tilde{n}^B \omega^{A(3)}_{B \nu}\\
    &=\partial^{(3)}_\nu \tilde{n}^A\nonumber\\
    &=0\,.\nonumber
\end{align}
The spatial covariant derivative can also be shown to inherit the nonmetricity property when this is also the case for the spacetime variant and that it similarly inherits metricity when not. 

Before continuing with the development of the $3+1$ formalism, it is useful to present the following relations as a consequence of nonmetricity. They are 
 \begin{align}
     n^\lambda \nabla_\rho n_\lambda & = \frac{1}{2} n^\lambda n^\nu Q_{\rho\lambda\nu}\,,\\
     n_\lambda \nabla_\rho n^\lambda & = - \frac{1}{2} n^\lambda n^\nu Q_{\rho\lambda\nu}\,.
 \end{align}
Defining the acceleration $a_\lambda$ as $n^\sigma \nabla_\sigma n_\lambda$,  we get
 \begin{align}
     a^\epsilon & = n^\sigma \nabla_\sigma n^\epsilon + n^\sigma n_\lambda {Q_\sigma}^{\lambda\epsilon}\,,\\
     n^\lambda a_\lambda & = \frac{1}{2}n^\sigma n^\lambda n^\epsilon Q_{\sigma\lambda\epsilon}\,.
 \end{align}
It is now possible to define the extrinsic curvature $k$ as 
\begin{align}
    k_{\alpha\beta}&=-\gamma^{\nu}_{\;\;\alpha}\gamma^{\mu}_{\;\;\beta}\nabla_{\nu}n_{\mu}\,.\label{eq:extrinsikdef}
\end{align}
From the definition of $k_{\alpha\beta}$, one can easily show that it is a purely spatial tensor. The extrinsic curvature can be thought  of as the change of direction of the normal vector $n$ as it is moved along the foliation $\Sigma$ \cite{baumgarte}. From this definition, the following alternate expression for the extrinsic curvature can be derived
\begin{align}
    k_{\nu\mu}&=-\nabla_{\nu}n_{\mu}-n_{\nu}a_{\mu}-\frac{1}{2}n^\rho n^\beta n_\mu {\gamma^\alpha}_\nu Q_{\alpha\rho\beta}\,.
\end{align}
It should be noted that the covector extrinsic curvature is defined as 
\begin{align*}
    k^{\alpha\beta}&=-\gamma^{\nu\alpha}\gamma^{\mu\beta}\nabla_{\nu}n_{\mu}\,,
\end{align*}
and that,  due to nonmetricity, it is 
\begin{align}
    \gamma^{\nu\alpha}\gamma^{\beta}_{\;\;\mu}\nabla_{\nu}n^{\mu} =-n_\lambda \gamma^{\alpha\nu}{\gamma^\beta}_\mu {Q_\nu}^{\lambda\mu} - k^{\alpha\beta}\,.
\end{align}

Having now defined all the basic building blocks of a generalized $3+1$ formalism, we consider the terms which can be used to characterize gravitation in such a theory. We define the Riemann and the torsion tensors in terms of the general affine connection $\Gamma^\rho_{\lambda\mu}$ as follows \cite{PereiraBook} and the nonmetricity term once again for convenience
\begin{align}
    R^{\;\rho}_{\;\;\lambda\nu\mu}&=\partial_\nu{\Gamma^\rho_{\lambda\mu}}-\partial_\mu{\Gamma^\rho_{\lambda\nu}}\,,\label{eq:Riemann_def}
    +\Gamma^\rho_{\alpha\nu}\Gamma^\alpha_{\lambda\mu}-\Gamma^{\rho}_{\alpha\mu}{\Gamma^\alpha_{\lambda\nu}}\,,\\
    T^{\lambda}_{\;\;\mu\nu}&=\Gamma^\lambda_{\nu\mu}-\Gamma^\lambda_{\mu\nu}\,,\label{eq:Torsion_def}\\
    Q_{\lambda\mu\nu} &=\nabla_\lambda g_{\mu\nu}\label{eq:Non-Metricity_def}\,,
\end{align}
where we note that, for a general affine connection, the Riemann tensor only possesses the antisymmetry $R^{\chi}_{\;\;\lambda [\rho\pi ]}$.

Through the definition of the torsion tensor it is also possible to show that it is related to the antisymmetry of the extrinsic curvature through
\begin{align}\label{eq:antisymmetric_extrinsic}
    k_{[\alpha\beta ]}&=\;\;\gamma^{\nu}_{\;\;\alpha}\gamma^{\mu}_{\;\;\beta}n_{\sigma}T^{\sigma}_{\;\;\nu\mu}\,.
\end{align}

Using the above definitions, the commutator of the spacetime covariant derivative of any vector $V^\lambda$ and covector $V_\lambda$ can be shown to be \cite{carroll2004spacetime}
\begin{align}
    \nabla_{[\nu}\nabla_{\mu]}V^{\lambda}=T^\sigma_{\;\;\mu\nu}\nabla_{\sigma}V^\lambda+V^{\sigma}R^\lambda_{\;\;\sigma\nu\mu}\,,\\
    \nabla_{[\nu}\nabla_{\mu]}V_{\lambda}=T^\sigma_{\;\;\mu\nu}\nabla_{\sigma}V_\lambda+V_{\sigma}R^\sigma_{\;\;\lambda\mu\nu}\,.
\end{align}
Here we note that the commutator is independent of nonmetricity.

Taking the commutator of the purely spatial covariant derivative applied to a general purely spatial vector $V^{\beta}_{(3)}$ and spatial co-vector $V_{\beta}^{(3)}$, one obtains the following equations
\begin{align}\label{eq:gen_antisymmetry_of_D}
    D_{[\sigma}D_{\alpha]}V^{\beta}&=T^{\lambda (3)}_{\;\;\alpha\sigma}D_{\lambda}V^{\beta}_{(3)} + V^{\lambda}_{(3)}R^{\beta (3)}_{\;\;\lambda\sigma\alpha}\\
    &=\;\gamma^{\rho}_{\;\;\sigma}\gamma^{\pi}_{\;\;\alpha}\gamma^{\beta}_{\;\;\chi}\left(T^{\lambda (4)}_{\;\;\pi\rho}D_{\lambda}V^{\chi}_{(3)}+ V^{\lambda}_{(3)}R^{\chi (4)}_{\;\;\lambda\rho\pi}\right)\nonumber\\ 
    &-V^{\lambda}_{(3)}k_{[\alpha | \lambda}{k_{| \sigma ]}}^{\beta} - \gamma^{\rho}_{\;\;[\sigma|}\gamma^{\beta}_{\;\;\chi}V^{\lambda}_{(3)} k_{|\alpha]\lambda}n_\mu {Q_\rho}^{\mu\chi}\,,\nonumber 
\end{align}
and
\begin{align}
    D_{[\sigma}D_{\alpha]}V_{\beta}&=T^{\lambda (3)}_{\;\;\alpha\sigma}D_{\lambda}V_{\beta}^{(3)} + V_{\lambda}^{(3)}R^{\lambda (3)}_{\;\;\beta\alpha\sigma}\\
    &=\;\gamma^{\rho}_{\;\;\sigma}\gamma^{\pi}_{\;\;\alpha}\gamma^{\chi}_{\;\;\beta}\left(T^{\lambda (4)}_{\;\;\pi\rho}D_{\lambda}V_{\chi}^{(3)}+ V_{\lambda}^{(3)}R^{\lambda (4)}_{\;\;\chi\pi\rho}\right)\nonumber\\ 
    &-V^{\lambda}_{(3)}k_{[\alpha | \lambda}{k_{| \sigma ] \beta}} - \gamma^{\pi}_{\;\;[\alpha|}n_\lambda V_{\nu}^{(3)} {Q_\pi}^{\lambda\nu} k_{|\sigma]\beta}\,.\nonumber 
\end{align}

Once a particular connection is chosen,  this equation will play a pivotal role in determining what is called the Gauss equation of the theoretical framework in question. The Gauss equation shows the relation between the four dimensional tensor used to characterize gravity like the Riemann or the torsion tensors and their three dimensional counterparts. With regards to the nonmetricity term $Q_{\alpha\mu\nu}$, it is not necessary for a connection, or theory, to be chosen to derive its Gauss equation. Starting from the spatial covariant derivative of the spacetime metric, we get
\begin{align}
    D_\rho g_{\mu\nu} &= \gamma^{\lambda}_{\;\;\rho}\gamma^{\beta}_{\;\;\mu}\gamma^{\alpha}_{\;\;\nu}\nabla_\lambda g_{\beta\alpha}= \gamma^{\lambda}_{\;\;\rho}\gamma^{\beta}_{\;\;\mu}\gamma^{\alpha}_{\;\;\nu}Q_{\lambda\beta\alpha}\\
    &=  \gamma^{\lambda}_{\;\;\rho}\gamma^{\beta}_{\;\;\mu}\gamma^{\alpha}_{\;\;\nu}\nabla_\lambda \left(\gamma_{\beta\alpha}-n_\beta n_\alpha\right)\nonumber\\
    &=D_\rho \gamma_{\mu\nu}\nonumber = Q_{\rho\mu\nu}^{\;(3)}\,.\nonumber
\end{align}
From this, we determine that, independent of the theory being considered, the Gauss equation for the nonmetricity term is
\begin{equation}\label{eq:non-metricity_Gauss}
    \gamma^{\lambda}_{\;\;\rho}\gamma^{\beta}_{\;\;\mu}\gamma^{\alpha}_{\;\;\nu}Q_{\lambda\beta\alpha} = Q_{\rho\mu\nu}^{\;(3)}\,.
\end{equation}

\subsection{Lie derivatives and generalized evolution equations}

To obtain the generalized evolution equations for the formalisms, we first define a general Lie derivative and explore some of its properties. For a general global 4-vector, $V^\nu$, and $4$-covector, $V_\nu$, the Lie derivative along a general vector $\chi^\mu$ is \cite{LieAsPartial1,LieAsPartial2,baumgarte}
\begin{align}
    \mathcal{L}_{\chi}V^\nu&=\chi^\mu\partial_\mu V^\nu -V^\mu\partial_\mu \chi^\nu\\
    &=\chi^\mu\nabla_\mu V^\nu -V^\mu\nabla_\mu \chi^\nu+V^\mu\chi^\lambda T^\nu_{\;\;\lambda\mu}\,,\nonumber\\
    \mathcal{L}_{\chi}V_\nu&=\chi^\mu\partial_\mu V_\nu +V_\mu\partial_\nu \chi^\mu\\
    &=\chi^\mu\nabla_\mu V_\nu +V_\mu\nabla_\nu \chi^\mu+V_\lambda \chi^\mu T^\lambda_{\;\;\mu\nu}\,.\nonumber
\end{align}
Visually this can be understood as the change in pointing direction and magnitude of the vector $V^\nu$ as it is infinitesimally dragged along the vector $\chi^\mu$ on the manifold.

In the case of a tensor with mixed global and local indices, a Lie derivative along a global vector is blind to the local indices and acts only on the global indices. This is understandable as locally no gravitational effect can be felt and there is nothing to evolve on the Minkowski metric. In a more general, purely geometric setting, going from one metric to another, it  is simply a transformation of the metric. The secondary background metric does not have to change when the primary changes, only the transformation or mapping tensor must change. In our case, the mapping tensor is the tetrad and as such it  evolves as the metric evolves while conserving the secondary metric. Given this fact, tensors with mixed indices will be treated as such
\begin{align}
    \mathcal{L}_{\chi}V^\nu_{\;\;A}&=\chi^\mu\partial_\mu V^\nu_{\;\;A} -V^\mu_{\;\;A}\partial_\mu \chi^\nu\,.
\end{align}
Let us now consider the Lie derivatives along the global normal $n_\mu$. We note that spatial-ness is conserved when such a derivative is applied to a general $3$-covector but not when applied to a general $3$-vector giving
\begin{align}
    n_\nu\mathcal{L}_{n}V^\nu_{(3)}&=-\frac{1}{\alpha}\partial_\mu(\alpha) V^\mu_{(3)}=\partial_\mu (Ln(\alpha^{-1}))V^\mu_{(3)}\,,\nonumber
\end{align}
The Lie derivative along the normal and its double action are used instead of any other derivative to obtain the evolution equations as they give the evolution of the fundamental variables of our formalisms, as they iterate through time. While a form of the evolution equations for a general affine connection can be obtained, it is necessary to choose a specific connection and Lagrangian, thus a theory, in order to obtain the final form of the equations that can actually be solved numerically to find the fundamental variables. The reason is that one of the final steps involves substituting the field equations of whatever theory one is considering into the evolution equations.

Noting that the Lie derivative of the  upper local index spatial tensor along any vector is spatial on that local index since
\begin{align}
    \bar{n}_A\mathcal{L}_{\chi}U^{A(3)}_{\;\;\nu} =& \bar{n}_A \chi^\mu \partial_\mu U^{A(3)}_{\;\;\nu} + \bar{n}_A U^{A(3)}_{\;\;\mu} \partial_\nu \chi^\mu\\
    =&-U^{A(3)}_{\;\;\nu}  \chi^\mu \partial_\mu \bar{n}_A\nonumber\\
    =&0\,.\nonumber
\end{align}
That being said, this is only true due to our choice of local metric, i.e., the Minkowski metric. As such, the Lie derivative of the tetrad is fully spatial. This result similarly applies for the second order Lie derivative of the tetrad and so on.

Thus, we present the first and second order Lie derivatives of the spatial metric and spatial tetrad along the normal vector for a general affine connection with nonmetricity
\begin{align}\label{eq:firstlieofmetric_gen}
\mathcal{L}_{n}\gamma_{\mu\nu}&= \gamma_{\sigma\left(\nu|\right.} \gamma^\beta_{\;\;\left.|\mu\right).} n^\lambda T^{\sigma}_{\;\;\lambda\beta} + 2\gamma^{\alpha}_{\;\;\nu} \gamma^{\beta}_{\;\;\mu}n_\lambda {L^{\lambda}}_{\alpha\beta}  - k_{\left(\mu\nu\right)}\\
&=A_{(\mu\nu)} + B_{(\mu\nu)} - k_{\left(\mu\nu\right)}\nonumber\,,
\end{align}

\begin{align}\label{eq:doublelieofmetric_Torsion}
\mathcal{L}_{n}A_{\mu\nu}&=-\gamma^{\alpha}_{\;\;\nu}
\gamma^{\beta}_{\;\;\mu}\nabla_\lambda {{T_\alpha}^\lambda}_{\beta}+D_\lambda {{T_\nu}^\lambda}_{\mu}\\
&+\gamma^{\alpha}_{\;\;\nu}
\gamma^{\beta}_{\;\;\mu} a_\sigma {{T_\alpha}^\sigma}_{\beta}+\gamma^{\beta}_{\;\;\mu} n_\sigma n^\epsilon {{T_\epsilon}^\sigma}_\beta \gamma^{\alpha}_{\;\;\nu}a_\alpha\nonumber \\
&-A_{\lambda\mu}\left(\gamma^{\lambda}_{\;\;\sigma}
\gamma^{\alpha}_{\;\;\nu} n_\epsilon {Q_\alpha}^{\epsilon\sigma}+{k_\nu}^{\lambda}\right) \nonumber \\
&-A_{\nu\lambda}\left(\gamma^{\lambda}_{\;\;\sigma}
\gamma^{\beta}_{\;\;\mu} n_\epsilon {Q_\beta}^{\epsilon\sigma}+{k_\mu}^{\lambda}\right) \nonumber \\
&+A_{\sigma\mu}{A^{\sigma}}_{\nu}+A_{\nu\sigma}{A^{\sigma}}_{\mu}\,,\nonumber 
\end{align}

\begin{align}\label{eq:doublelieofmetric_nonmeticity}
\mathcal{L}_{n}B_{\mu\nu}&=-\gamma^{\alpha}_{\;\;\nu}
\gamma^{\beta}_{\;\;\mu}\nabla_\lambda {L^\lambda}_{\alpha\beta}+D_\lambda {L^\lambda}_{\mu\nu}\\
&+\gamma^{\alpha}_{\;\;\nu}
\gamma^{\beta}_{\;\;\mu}\left( a_\sigma {L^\sigma}_{\alpha\beta} 
+a_{\left(\beta\right|}{L^\sigma}_{\epsilon\left|\alpha\right)}n_{\sigma}n^{\epsilon}\right)\nonumber \\
&-\gamma^{\epsilon}_{\;\;\left(\nu\right|}
\gamma^{\chi}_{\;\;\left|\mu\right)}B_{\alpha\chi}\left(n_\sigma{Q_\epsilon}^{\sigma\alpha}+{k_\epsilon}^{\alpha}\right) \nonumber \\
&+B_{\sigma\left(\mu\right|}{A^{\sigma}}_{\left|\nu\right)}\,,\nonumber 
\end{align}

\begin{align}\label{eq:doublelieofmetric_Curvature}
\mathcal{L}_{n}k_{\mu\nu}&=\gamma^{\alpha}_{\;\;\nu}
\gamma^{\beta}_{\;\;\mu} n^\lambda \nabla_\lambda \nabla_\alpha n_\beta\\
&+\gamma^{\alpha}_{\;\;\nu}
\gamma^{\beta}_{\;\;\mu} \left(\frac{1}{2}n^\chi n^\epsilon Q_{\alpha\chi\epsilon} a_\beta +a_\beta a_\alpha \right)\nonumber \\
&-k_{\lambda\mu}\left(\gamma^{\lambda}_{\;\;\sigma}
\gamma^{\alpha}_{\;\;\nu} n_\epsilon {Q_\alpha}^{\epsilon\sigma}+{k_\nu}^{\lambda}\right) \nonumber \\
&-k_{\nu\lambda}\left(\gamma^{\lambda}_{\;\;\sigma}
\gamma^{\beta}_{\;\;\mu} n_\epsilon {Q_\beta}^{\epsilon\sigma}+{k_\mu}^{\lambda}\right) \nonumber \\
&+k_{\sigma\mu}{A^{\sigma}}_{\nu}+k_{\nu\sigma}{A^{\sigma}}_{\mu}\,,\nonumber 
\end{align}

\begin{align}\label{eq:gen_tetrad_first_lie}
\mathcal{L}_{n}e^{A(3)}_{\;\;\nu}&=n^\mu \partial_\mu e^{A(3)}_{\;\;\nu} + e^{A(3)}_{\;\;\mu} \partial_\nu n^\mu\\
&=n^\mu \nabla_\mu e^{A(3)}_{\;\;\nu} + e^{A(3)}_{\;\;\mu} \nabla_\nu n^\mu \nonumber \\
&\;\;\;\;+ n^\mu e^{A(3)}_{\;\;\lambda} \left(\Gamma^\lambda_{\nu\mu}  -  \Gamma^\lambda_{\mu\nu}\right)- n^\mu e^{B(3)}_{\;\;\nu}  \omega^{A}_{\;\;B\mu}\nonumber\\
&= {\tilde{\gamma}^{A}}_C{\gamma^{\alpha}}_\nu n^\mu\nabla_\mu {e^C}_\alpha- e^{A(3)}_{\;\;\mu} n_\epsilon {{\gamma}^{\mu}}_\sigma {\gamma^{\alpha}}{Q_\alpha}^{\epsilon\sigma}\nonumber\\ 
&\;\;\;\;-e^{A(3)}_{\;\;\mu}\gamma_{\alpha\nu}k^{\alpha\mu}+e^{A(3)}_{\;\;\lambda} {\gamma^\lambda}_\epsilon {\gamma^{\alpha}}_\nu n^\mu  T^{\epsilon}_{\;\;\mu\alpha}\nonumber\\
&\;\;\;\;-  e^{F(3)}_{\;\;\nu} {\tilde{\gamma}^{B}}_F {\tilde{\gamma}^{A}}_C n^\mu \omega^{C}_{\;\;B\mu},\nonumber\nonumber\\
&={\mathcal{Q}^{A}}_\nu - {\mathcal{B}^{A}}_\nu - {k_{\nu}}^A + {A^{A}}_{\nu} - {\mathcal{C}^A}_\nu\,, \nonumber
\end{align}

\begin{align}\label{eq:gen_tetrad_second_lie}
\mathcal{L}_{n}U^{A}_{\;\;\nu}&= {\tilde{\gamma}^{A}}_C{\gamma^{\alpha}}_\nu n^\lambda \left(\nabla_\lambda {U^C}_\alpha -\nabla_\alpha {U^C}_\lambda \right) \\
&\;\;\;\;- {U^B}_\nu {\mathcal{C}^A}_B + {U^A}_\lambda {A^\lambda}_\nu\,, \nonumber
\end{align}
where ${{U}^A}_\nu$ is a general spatial tensor that can be substituted by each of the terms in the first Lie derivative of the tetrad to obtain a final general form for the second Lie derivative of the tetrad similar to that of the metric. 

We end this section by presenting the Lie derivative of a general spatial tensor along $\alpha n^\mu$. It can be shown that, even for a general connection, the Lie derivative along $\alpha n^\mu$ of such a tensor is itself spatial. Taking this Lie derivative of some tensor ${X^{\epsilon_1 ... \epsilon_i}}_{\sigma_1 ... \sigma_j}$, where $i$ and $j$ are integers, it results in
\begin{align}\label{eq:lie_an}
    \mathcal{L}_{\alpha n}{X^{\epsilon_1 ... \epsilon_i}}_{\sigma_1 ... \sigma_j}=& \alpha n^\lambda \partial_\lambda {X^{\epsilon_1 ... \epsilon_i}}_{\sigma_1 ... \sigma_j}\\
    &\;\;- \sum_{p=1}^{i} {X^{\epsilon_1 ..\lambda_p.. \epsilon_i}}_{\sigma_1 ... \sigma_j} \partial_{\lambda_p} \left(\alpha n^{\epsilon_p} \right)\nonumber\\
    &\;\;\;\;+ \sum_{q=1}^{j} {X^{\epsilon_1 .. \epsilon_i}}_{\sigma_1 ..\lambda_q.. \sigma_j} \partial_{\sigma_q} \left(\alpha n^{\lambda_q} \right)\,.\nonumber
\end{align}
Thus, taking $n^{\sigma_k}\mathcal{L}_{n}{X^{\epsilon_1 ... \epsilon_i}}_{\sigma_1 ... \sigma_j}$ for $1\leq k \leq j$ will result in the second term of Eq.~(\ref{eq:lie_an}) and all of the third term except for when $k=q$ to become zero. What  remains of Eq.~(\ref{eq:lie_an}) is 
\begin{align}
    n^{\sigma_k}\mathcal{L}_{\alpha n}{X^{\epsilon_1 ... \epsilon_i}}_{\sigma_1 ... \sigma_j}=& \alpha n^{\sigma_k} n^\lambda \partial_\lambda {X^{\epsilon_1 ... \epsilon_i}}_{\sigma_1 ... \sigma_j}\\
    &\;\;\;\;+ n^{\sigma_k} {X^{\epsilon_1 .. \epsilon_i}}_{\sigma_1 ..\lambda.. \sigma_j} \partial_{\sigma_k} \left(\alpha n^{\lambda} \right)\nonumber\\
    =&\alpha n^{\sigma_k} n^\lambda \left(\partial_\lambda {X^{\epsilon_1 ... \epsilon_i}}_{\sigma_1 ..\sigma_k.. \sigma_j}\right.\nonumber\\
    &\;\;\;\;\left.-  \partial_{\sigma_k}{X^{\epsilon_1 .. \epsilon_i}}_{\sigma_1 ..\lambda.. \sigma_j}  \right)\nonumber\\
    =&0\,.\nonumber
\end{align}
Taking $n_{\epsilon_k}\mathcal{L}_{n}{X^{\epsilon_1 ... \epsilon_i}}_{\sigma_1 ... \sigma_j}$ for $1\leq k \leq i$, it will result in the third term of Eq.~(\ref{eq:lie_an}) and all of the second term except for when $k=p$ to become zero. what remains of Eq.~(\ref{eq:lie_an}) is 
\begin{align}
    n_{\epsilon_k} \mathcal{L}_{\alpha n}{X^{\epsilon_1 ... \epsilon_i}}_{\sigma_1 ... \sigma_j}=& \alpha n_{\epsilon_k} n^\lambda \partial_\lambda {X^{\epsilon_1 ... \epsilon_i}}_{\sigma_1 ... \sigma_j}\\
    &-n_{\epsilon_k}{X^{\epsilon_1 ..\lambda.. \epsilon_i}}_{\sigma_1 ... \sigma_j} \partial_{\lambda} \left(\alpha n^{\epsilon_k} \right)\nonumber\\
   =& n_{\epsilon_k} \partial_\lambda \left(\alpha\right)\left( n^{\epsilon_k}{X^{\epsilon_1 ..\lambda.. \epsilon_i}}_{\sigma_1 ... \sigma_j}  \right.\nonumber\\
    &\left.  -n^\lambda {X^{\epsilon_1 ... \epsilon_i}}_{\sigma_1 ... \sigma_j} \right)\nonumber\\
    &+{X^{\epsilon_1 ..\lambda.. \epsilon_i}}_{\sigma_1 ... \sigma_j} \partial_{\lambda} \left( \alpha \right)\nonumber\\
    =&0\,,\nonumber
\end{align}
where 
\begin{align}
    \alpha\partial_\lambda \left(n_{\epsilon_k}\right) =& -\alpha\partial_\lambda \left(\alpha\partial_{\epsilon_k}\left\{t\right\}\right)\\
    =& n_{\epsilon_k} \partial_\lambda \left(\alpha\right)\,,\nonumber    
\end{align}
was used.

\section{Gauss Constraint and Evolution equations in Specific theories}\label{sec:specifics}

In this section, the generalized $3+1$ metric and tetrad formalisms set up in Sec.~\ref{Sec:General_for} will be applied to three actual theories of interest by choosing their respective connections. In the first case, the Levi-Civita connection is chosen specifying GR as the theory under consideration.

\subsection{Gauss constraint and evolution equations in general relativity}\label{sec:gr_decom}

GR is built on the affine Levi-Civita connection which can be defined through the $4$-metric as \cite{carroll2004spacetime}
\begin{align}\label{eq:levi-civita}
    \lc{\Gamma}^{\sigma}_{\mu\nu}&:=\frac{1}{2}g^{\sigma\rho}\left(\partial_\mu g_{\rho\nu}+\partial_\nu g_{\nu\rho}-\partial_\rho g_{\mu\nu}\right)\,.
\end{align}
From this definition,  one can easily determine that this connection is symmetric in the bottom two indices. It can also be shown that this connection is not a tensor as indicated in the placement of the indices in Eq.~(\ref{eq:levi-civita}). It is this connection which characterizes curvature rather than the metric itself.

It should be noted that, unless otherwise specified, throughout this subsection all covariant derivatives, Lie derivatives and tensors are derived through the Levi-Civita connection. 

Choosing to work with the Levi-Civita connection, it is important to point out some natural consequences of the symmetry of this connection both in general and more specifically to our $3+1$ decomposition approach. The Riemann tensor, as defined in Eq.~(\ref{eq:Riemann_def}), acquires two additional symmetries apart from the antisymmetry in the last two indices. These are given below for ease of reference \cite{carroll2004spacetime, Boskoff}
\begin{align}
    \lc{R}_{\rho\sigma\mu\nu}=-\lc{R}_{\sigma\rho\mu\nu}\,,\\
    \lc{R}_{\rho\sigma\mu\nu}=-\lc{R}_{\mu\nu\sigma\rho}\,.
\end{align} 
The Riemann tensor also acquires the following cyclic property
\begin{equation}
    \lc{R}_{\rho\sigma\mu\nu}+\lc{R}_{\rho\mu\nu\sigma}+\lc{R}_{\rho\nu\sigma\mu}=0\,.
\end{equation}
Other major consequences are that the torsion tensor, by definition, is now a zero tensor and  the theory assumes metricity so all nonmetricity and  disformation tensors are zero. This leads to another important result which is that the definition of the Lie derivative can be equivalently written using partial derivatives and covariant derivatives for purely global tensors. This is the case as the difference between using a partial and a covariant derivative consists only of torsion tensor terms. Thus, in GR, when taking the Lie derivative of any global tensor along any vector, the following definitions are equivalent 
\begin{align}
    \mathcal{L}_{\chi}V^\nu_{\;\;a\mu}&=\chi^\sigma\partial_\sigma V^\nu_{\;\;a\mu} -V^\sigma_{\;\;a\mu}\partial_\sigma \chi^\nu+V^\nu_{\;\;a\sigma}\partial_\mu \chi^\sigma\\
    &=\chi^\sigma\lc{\nabla}_\sigma V^\nu_{\;\;a\mu} -V^\sigma_{\;\;a\mu}\lc{\nabla}_\sigma \chi^\nu+V^\nu_{\;\;a\sigma}\lc{\nabla}_\mu \chi^\sigma\,.\nonumber
\end{align} 
In the case of a tensor with local indices, this does not apply as spin connection terms will still  persist. A zero torsion tensor also results in a completely symmetric extrinsic curvature. Finally, it can be shown that the covariant derivative of the one-form $\Omega_\nu=\nabla_\nu t$ is zero as it is also composed by just a torsion term \cite{baumgarte}.

Having now defined the basic consequences of the symmetry of  Levi-Civita connection, we derive the Gauss-Codazzi constraint and the evolution equations for a $3+1$ formalism in GR based on the metric \cite{baumgarte}.

Given the generality of the $3$-vector $V^\lambda$, we retrieve the Riemann Gauss equation from the generalized relation in Eq.~\eqref{eq:gen_antisymmetry_of_D} between the antisymmetry of the purely spatial and space time covariant derivatives \cite{baumgarte}
\begin{align}\label{eq:reimann_Gauss}
    \lc{R}^{\beta (3)}_{\;\;\lambda\sigma\alpha}=\;\gamma^{\rho}_{\;\;\sigma}\gamma^{\pi}_{\;\;\alpha}\gamma^\nu_{\;\;\lambda}\gamma^{\beta}_{\;\;\chi}\lc{R}^{\chi}_{\;\;\nu\rho\pi}-\lc{k}_{[\alpha | \lambda}{\lc{k}_{| \sigma ]}}{}^{\beta}\,.
\end{align}
This equation relates the fully spatial Riemann tensor to its $4$-dimensional counterpart. Contracting once and then twice,  we obtain the equivalent relation for the Ricci tensor and the Ricci scalar
\begin{equation}\label{eq:Ricci_T_Gauss}
    \lc{R}^{(3)}_{\lambda\alpha}=\;\gamma^{\pi}_{\;\;\alpha}\gamma^\nu_{\;\;\lambda}\gamma^{\rho}_{\;\;\chi}\lc{R}^{\chi}_{\;\;\nu\rho\pi}-\lc{k}_{\alpha  \lambda}\lc{k}+\lc{k}_{\beta\lambda}{\lc{k}_{\alpha}}{}^{\beta}\,,
\end{equation}

\begin{equation}\label{eq:Ricci_S_Gauss}
    \lc{R}^{(3)}=\;\gamma^{\pi\nu}\gamma^{\rho}_{\;\;\chi}\lc{R}^{\chi}_{\;\;\nu\rho\pi}-\lc{k}^2+\lc{k}_{\beta\lambda}\lc{k}{}^{\lambda\beta}\,.
\end{equation}
Finally the Codazzi equation can also be shown to be given by 
\begin{align}\label{eq:Codazzi_GR}
    \lc{D}_{\left[{\mu}\right.} \lc{k}_{\left.\nu\right]\lambda}&=\;\gamma^{\rho}_{\;\;\nu}\gamma^{\pi}_{\;\;\mu}\gamma^{\beta}_{\;\;\lambda}n^{\sigma} \lc{R}_{\rho\pi\beta\sigma}\,.
\end{align}
We note that these equations fully agree  with those obtained in literature \cite{baumgarte,gourgoulhon}.

Another consequence of the Levi-Civita connection is that by putting the spin connection as subject of the formula of the covariant derivative of the tetrad, and applying the tetrad metricity property and substituting Eq.~\eqref{eq:levi-civita},  we can write the spin connection purely in terms of tetrads and the Minkowski metric, that is 
\begin{align}
    \omega^A_{\;\;C\nu}=&\frac{1}{2} \left(\bar{e}^{A\alpha} e^{B}_{\;\;\nu} e_{C}^{\;\;\beta} \left(\partial_\beta \bar{e}_{B\alpha}-\partial_\alpha \bar{e}_{B\beta}\right)\right.\nonumber\\
    &\left.+e_{C}^{\;\;\alpha} \left(\partial_\alpha e^{A}_{\;\;\nu}-\partial_\nu e^A_{\;\;\alpha}\right)+\bar{e}^{A\alpha} \left(\partial_\nu \bar{e}_{C\alpha}-\partial_\alpha \bar{e}_{C\nu}\right)\right)\,.\label{eq:spin_tet_def}
\end{align}
It should be noted that, for the sake of simplicity, terms like $\eta^{AB}e_B^{\;\;\alpha}$ are written as $\bar{e}^{A\alpha}$. This would be useful if a tetrad $3+1$ is considered for GR. An interesting by-product of this form of  spin connection is that the antisymmetry in the first two indices of the connection can be very clearly observed.

Finally, we consider the Ricci equation. This equation is understood to be the second order Lie derivative of the fundamental variable.  The $3+1$ formalism is being based on. Otherwise one may look at it as an extra evolution equation of some resulting variable from the first order Lie derivative of the fundamental variable for which an expression consisting only of such a fundamental variable is not know. This is effectively the concept of requiring two equations for two unknowns. 

Through simplifying the general form of this equation for the metric formalism, Eqs.~(\ref{eq:doublelieofmetric_Torsion}-\ref{eq:doublelieofmetric_Curvature}), using the Levi-Civita properties, assuming metricity and substituting in the Einstein field equations, we find
\begin{align}
    \mathcal{L}_{n}\gamma_{\mu\nu}&=-2 \lc{k}_{\mu\nu}\,, \\
    \mathcal{L}_{n}\lc{k}_{\mu\nu}&=-\frac{1}{2}\mathcal{L}_{n}\mathcal{L}_{n}\gamma_{\mu\nu}\label{eq:gr_so_metric_lie}\\
    &=\;n^\rho \gamma^{\alpha}_{\;\;\nu} \gamma^{\beta}_{\;\;\mu} n_{\chi}\lc{R}^\chi_{\;\;\alpha\rho\beta} - \lc{D}_{\mu} \lc{a}_{\nu} - \lc{a}_\mu \lc{a}_\nu - {\lc{k}_{\mu}}{}^\rho \lc{k}_{\nu\rho}\nonumber\\
    &=\;n^\rho \gamma^{\alpha}_{\;\;\nu} \gamma^{\beta}_{\;\;\mu} n_{\chi}\lc{R}^\chi_{\;\;\alpha\rho\beta} - \frac{1}{\alpha}\lc{D}_\mu \lc{D}_\nu \alpha - {\lc{k}_{\mu}}{}^\rho \lc{k}_{\nu\rho}\,,\nonumber
\end{align}
where $\lc{D}_{\mu} \lc{a}_{\nu} + \lc{a}_\mu \lc{a}_\nu$ can be shown to be equivalent to $\frac{1}{\alpha}\lc{D}_\mu \lc{D}_\nu\alpha$ given that the torsion tensor is zero.

Finally, the momentum and Hamiltonian constraints are derived to be
\begin{align}
    \lc{D}_{\lambda} \lc{k} - \lc{D}_{\chi} {\lc{k}_{\lambda}}^{\chi} &= 8\pi S_\lambda\,,\label{eq:gr_momentum_constraint}\\
    \lc{R}^{(3)}+\lc{k}^2-\lc{k}_{\beta\lambda}\lc{k}^{\lambda\beta} &= 16 \pi \rho\,.\label{eq:gr_hamiltonian_constraint}
\end{align} 
It should be noted that the metric evolution equations and the constraint equation agree perfectly with those found in literature \cite{baumgarte,gourgoulhon}.

\subsection{Gauss constraint and evolution equations in the teleparallel equivalent of general relativity}\label{sec:tegr_decom}

Here we consider TG, specifically the TEGR and perform a $3+1$ decomposition in the Weitzenb\"{o}ck gauge. In this case, all curvature terms are identically zero and metricity is assumed. Similarly to what was done in the case of GR, a tetrad formulation of  $3+1$ decomposition will be derived through an investigation starting with the  connection in this approach to gravity. In TG, the teleparallel connection defined in Eq.~\eqref{tetrad_defs} is used in terms of the $4$-tetrad and the spin connection
\begin{equation}
    \tg{\Gamma}^{\lambda}_{\mu\nu} := {e_{A}}^\lambda \partial_\nu {e^{A}}_\mu + {e_{A}}^\lambda {e^{B}}_\mu \omega^{a}_{B\nu}\,.
\end{equation}
Throughout this section,  all covariant derivatives, Lie derivatives and tensors are derived through the Weitzenb\"{o}ck connection (which is the teleparallel connection in the Weitzenb\"{o}ck gauge) and as such they are denoted by a hat accent, such as in $\tg{A}$. It is important to point out that even spatial components are denoted in this way.

Unlike in the GR case, the TG spin connection  has not an explicit expression in terms of fundamental variables. For this reason, we only consider tetrads in their Weitzenb\"{o}ck gauge, simplifying the resulting expressions. In the Weitzenb\"{o}ck gauge, the $4$-covariant derivative of the normal vector is zero resulting in a zero covariant derivative of the $3$-metric and the $3$-tetrad, in a zero extrinsic curvature tensor and in a zero acceleration vector.

After applying all of these results to Eq.~(\ref{eq:gen_antisymmetry_of_D}), we obtain the following equation
\begin{align}
    \tg{D}_{\lambda}V^{\beta}_{(3)}\left(\tg{T}^{\lambda (3)}_{\;\;\alpha\sigma}-\gamma^{\rho}_{\;\;\sigma}\gamma^{\pi}_{\;\;\alpha}\gamma^{\lambda}_{\;\;\nu}\tg{T}^{\nu (4)}_{\;\;\pi\rho}\right)=0\,.\label{eq:TEGR_Spin_0_antisymmetry_of_D}
\end{align}

There are three cases which would  result in this dot product being zero. The first is that $\tg{D}_{\lambda}V^{\beta}_{(3)}=0$, which is not possible since $V^{\beta}_{(3)}$ is a general $3$-vector. The second is that $\left(\tg{T}^{\lambda (3)}_{\;\;\alpha\sigma}-\gamma^{\rho}_{\;\;\sigma}\gamma^{\pi}_{\;\;\alpha}\gamma^{\lambda}_{\;\;\nu}\tg{T}^{\nu (4)}_{\;\;\pi\rho}\right)$ is perpendicular to $\tg{D}_{\lambda}V^{\beta}_{(3)}$ for all $3$-vectors $V^{\beta}_{(3)}$. However, the bracket term is in no-way dependent on the general $3$-vector $V^{\beta}_{(3)}$, then the bracket term would have to be purely temporal. This is a contradiction since applying a dot product on any of the free indices of the bracket term with a normal vector would result in zero
\begin{align}
    n_\lambda\left(\tg{T}^{\lambda (3)}_{\;\;\alpha\sigma}-\gamma^{\rho}_{\;\;\sigma}\gamma^{\pi}_{\;\;\alpha}\gamma^{\lambda}_{\;\;\nu}\tg{T}^{\nu (4)}_{\;\;\pi\rho}\right)&=0\,,\\\nonumber
    n^\alpha\left(\tg{T}^{\lambda (3)}_{\;\;\alpha\sigma}-\gamma^{\rho}_{\;\;\sigma}\gamma^{\pi}_{\;\;\alpha}\gamma^{\lambda}_{\;\;\nu}\tg{T}^{\nu (4)}_{\;\;\pi\rho}\right)&=0\,,\\\nonumber
    n^\sigma\left(\tg{T}^{\lambda (3)}_{\;\;\alpha\sigma}-\gamma^{\rho}_{\;\;\sigma}\gamma^{\pi}_{\;\;\alpha}\gamma^{\lambda}_{\;\;\nu}\tg{T}^{\nu (4)}_{\;\;\pi\rho}\right)&=0\,,\nonumber
\end{align}
implying that the bracket term is purely spatial. The only option left is that $\left(\tg{T}^{\lambda (3)}_{\;\;\alpha\sigma}-\gamma^{\rho}_{\;\;\sigma}\gamma^{\pi}_{\;\;\alpha}\gamma^{\lambda}_{\;\;\nu}\tg{T}^{\nu (4)}_{\;\;\pi\rho}\right)$ is in fact zero itself producing the  Gauss equation for torsion
\begin{equation}
    \tg{T}^{\lambda (3)}_{\;\;\alpha\sigma}=\gamma^{\rho}_{\;\;\sigma}\gamma^{\pi}_{\;\;\alpha}\gamma^{\lambda}_{\;\;\nu}\tg{T}^{\nu (4)}_{\;\;\pi\rho}\,.\label{eq:torsion_Gauss}
\end{equation}
Through this relation, similar relationships can be obtained for the torsion vector and scalar
\begin{align}
    \tg{T}^{(3)}_\alpha&=\gamma^{\rho}_{\;\;\lambda}\gamma^{\pi}_{\;\;\alpha}\gamma^{\lambda}_{\;\;\nu}\tg{T}^{\nu (4)}_{\;\;\pi\rho}\nonumber\\
    &=\tg{T}^{(4)}_\alpha+n_{\alpha}n^{\pi}\tg{T}^{(4)}_\pi+n_{\nu}n^{\rho}\tg{T}^{\nu (4)}_{\;\;\alpha\rho}\nonumber\\
    &=\gamma^{\lambda}_{\;\;\alpha}\tg{T}^{(4)}_\lambda+n_{\nu}n^{\rho}\tg{T}^{\nu (4)}_{\;\;\alpha\rho}\,,\label{eq:tor_vec_gauss}
\end{align}
\begin{align}
    \tg{T}^{(3)}=&\tfrac{1}{4} \tg{T}_{\lambda }^{(3)\mu \nu } \tg{T}^{\lambda (3)}_{\;\;\mu \nu } + \tfrac{1}{2} \tg{T}^{\lambda(3)}_{\;\;\mu \nu } \tg{T}^{\nu \mu(3) }_{\;\;\;\;\lambda } - \tg{T}^{\lambda(3) }_{\;\;\mu \lambda } \tg{T}^{\nu \mu (3)}_{\;\;\;\;\nu }\nonumber\\
    =&\tg{T}^{(4)}+2 n^{\lambda } n^{\mu } \tg{T}^{\alpha }{}_{\nu \alpha } \tg{T}_{\lambda \mu }{}^{\nu } + \tfrac{1}{4} n^{\lambda } n^{\mu } \tg{T}_{\lambda }{}^{\nu \alpha } \tg{T}_{\mu \nu \alpha }\nonumber\\
    &+ n^{\lambda } n^{\mu } \tg{T}_{\lambda }{}^{\nu \alpha } \tg{T}_{\nu \mu \alpha } + \tfrac{1}{2} n^{\lambda } n^{\mu } \tg{T}_{\alpha \mu \nu } \tg{T}^{\nu }{}_{\lambda }{}^{\alpha }\nonumber \\
    &+ \tfrac{1}{2} n^{\lambda } n^{\mu } \tg{T}_{\nu \mu \alpha } \tg{T}^{\nu }{}_{\lambda }{}^{\alpha } - n^{\lambda } n^{\mu } \tg{T}^{\alpha }{}_{\mu \alpha } \tg{T}^{\nu }{}_{\lambda \nu }\nonumber\\
    =&\tg{T}^{(4)} +\frac{2}{\alpha}\tg{T}^{\nu(3)}\partial^{(3)}_{\nu}\left(\alpha\right)+\frac{1}{2}\tg{A}^\nu_\lambda\left(\tg{A}_\nu^\lambda+\tg{A}^\lambda_\nu\right)-\tg{A}^2\,,\label{eq:tor_scal_gauss}
\end{align}
where $A$ is the first order Lie derivative of the spatial tetrad along the normal vector to the foliations $n$.

We now move on to the evolution equations by considering the TEGR field equations \cite{fieldequasions,Krssak:2015oua} as given in Eq.~(\ref{TEGR_FEs}). After setting the Weitzenb\"{o}ck gauge and some minor restructuring, these field equations can be written as 
\begin{align}
    \tg{S}_{\alpha  \sigma  }{}^{\rho  } \tg{T}^{\lambda  }{}_{\rho  \lambda  } - \tg{S}^{\rho  \lambda  }{}_{\sigma  }&\tg{T}_{\rho  \lambda  \alpha  } - \tfrac{1}{2} \tg{S}_{\alpha  }{}^{\rho  \lambda  } \tg{T}_{\sigma  \lambda  \rho  }\nonumber\\
    &+ \tfrac{1}{2} g_{\alpha  \sigma  } \tg{T} - \tg{\nabla}^{\lambda  }\tg{S}_{\alpha  \lambda  \sigma  }=\Theta_{\alpha \sigma}\,.\label{EE_FE}
\end{align}
Contracting these equations, we obtain an alternative expression for the torsion scalar in terms of the energy-momentum scalar
\begin{align}
    \tg{T} &= \Theta +2 \tg{T}^{\alpha }{}_{\rho \alpha } \tg{T}^{\lambda }{}_{\lambda }{}^{\rho }  + 2 \tg{\nabla}_{\rho }\tg{T}^{\lambda }{}_{\lambda }{}^{\rho }\,.\label{Energy_Momentum_Scalar}
\end{align}
The generalized tetrad evolution equations in Eq.(\ref{eq:gen_tetrad_first_lie}) and Eq.(\ref{eq:gen_tetrad_second_lie}) then reduce to
\begin{align}\label{eq:TEGR_Spin0_tetrad_both_lie}
    \mathcal{L}_{n}e^{A(3)}_{\;\;\nu}&=n^\lambda e^{A(3)}_{\;\;\rho} \gamma^\sigma_{\;\;\nu} \tg{T}^{\rho}_{\;\;\lambda\sigma}\nonumber\\
    &=\tg{A}^A_{\;\;\nu}\,,\\
    \mathcal{L}_{n}\tg{A}^A_{\;\;\nu}&=n^\rho n^\mu e^{A(3)}_{\;\;\lambda}\gamma^\alpha_{\;\;\nu}\left(\tg{\nabla}_\rho \tg{T}^{\lambda}_{\;\;\mu\alpha} + \tg{T}^{\lambda}_{\;\;\mu\pi}\tg{T}^{\pi}_{\;\;\rho\alpha}\right)\nonumber\\
    &=\tg{D}_\rho \tg{T}^{A\rho}_{(3)\nu}+\tg{A}^A_{\;\;\rho}\tg{A}^\rho_{\;\;\nu}+e^{A(3)}_{\;\;\sigma}\gamma^{\sigma}_{\;\;\lambda}\gamma^{\alpha}_{\;\;\nu}\tg{\nabla}_\rho \tg{T}^{\lambda\;\;\rho}_{\;\;\alpha}\,.
\end{align}
In order to find the final from of the TEGR evolution equations, we now substitute the field equations into the evolution equations
\begin{align}\label{eq:TEGR_Spin0_tetrad_both_lie_Final}
    \mathcal{L}_{n}e^{A(3)}_{\;\;\nu}&=n^\lambda e^{A(3)}_{\;\;\rho} \gamma^\sigma_{\;\;\nu} \tg{T}^{\rho}_{\;\;\lambda\sigma}\nonumber\\
    &=\tg{A}^A_{\;\;\nu}\,,\\
    \mathcal{L}_{n}\tg{A}^a_{\;\;\nu}&=\tg{D}_\rho \tg{T}^{A\rho}_{(3)\nu}+\tg{A}^A_{\;\;\rho}\tg{A}^\rho_{\;\;\nu}+e^{A(3)}_{\;\;\sigma}\left[{\tg{A}_\nu}^\rho {\tg{A}^\sigma}_\rho\right.\nonumber\\
    &-{\tg{A}^\rho}_\nu {\tg{A}_\rho}^\sigma - \frac{1}{2} \tg{T}_{\;\;\nu}^{(3)\rho\chi}\tg{T}^{\sigma(3)}_{\;\;\rho\chi}+\tg{T}^{\rho(3)\chi}_{\;\;\;\;\nu}\left(\tg{T}_{\rho\;\;\chi}^{\;\;\sigma(3)}+\tg{T}_{\chi\;\;\rho}^{\;\;\sigma(3)}\right)\nonumber\\
    &+\frac{2}{\alpha^2}\partial_\nu^{(3)}(\alpha)\partial^\sigma_{(3)}(\alpha)-\tg{T}^{\chi(3)}_{\;\;\rho\chi}\left(\tg{T}_{\;\;\nu}^{(3)\sigma\rho}+\tg{T}^{\sigma(3)\rho}_{\;\;\;\;\nu}\right)\nonumber\\
    &-\frac{1}{\alpha}\partial_\rho^{(3)}(\alpha)\left(\tg{T}_{\;\;\nu}^{(3)\sigma\rho}+\tg{T}^{\sigma(3)\rho}_{\;\;\;\;\nu}\right)-\tg{A}\left( {\tg{A}_\nu}^\sigma+{\tg{A}^\sigma}_\nu\right)\nonumber\\
    &+\tg{D}_\nu \tg{T}^{\rho\sigma}_{(3)\rho}+\tg{D}^\sigma \tg{T}^{\rho(3)}_{\;\;\nu\rho}+\tg{D}_\nu\left(\frac{1}{\alpha}\partial^\sigma_{(3)}\{\alpha\}\right)\nonumber\\
    &+{\tg{D}^\sigma}\left(\frac{1}{\alpha}\partial^{(3)}_\nu\{\alpha\}\right)+8\pi G\left(2 {S_\nu}^\sigma - {\gamma^\sigma}_\nu\left\{S-\rho\right\} \right)\left.\right]\nonumber\\
    &-e^{A(3)}_{\;\;\sigma}\gamma^{\sigma}_{\;\;\lambda}\gamma^{\alpha}_{\;\;\nu}\tg{\nabla}_\rho \tg{T}_{\alpha}^{\;\;\lambda\rho}\,,
\end{align}
where ${S_\nu}^\sigma$ and $S$ are the spatial stress and scalar stress while  $\rho$ is the density. At this point, we note an important issue with this $3+1$ decomposition of TEGR. Specifically we note that the final term $e^{A(3)}_{\;\;\sigma}\gamma^{\sigma}_{\;\;\lambda}\gamma^{\alpha}_{\;\;\nu}\tg{\nabla}_\rho \tg{T}_{\alpha}^{\;\;\lambda\rho}$ cannot be converted into a purely spatial form which is necessary to obtain a valid form for the evolution equations. Such terms are normally substituted out through the field equations, however, due to the symmetry of the field equations and asymmetry of the second evolution equation, this term is not eliminated. If instead of considering the tetrad evolution equations, one considers the metric ones, the problem does not feature due to the symmetry of the metric. However, this creates a new issue since it is not possible to write torsion tensors purely in terms of metrics making the equations insoluble. For completeness, we finalize the remainder of the $3+1$ decomposition.

An interesting relationship between GR and TEGR comes out from manipulating the definition of the first Lie derivative of the spatial tetrad or spatial metric. Taking the symmetry of $\tg{A}^A_{\;\;\nu}$, one gets the following
\begin{align}\label{eq:A=k}
    \tg{A}^{A}_{\;\;\nu}+\tg{A}_{\nu}^{\;\;A}=&n^\lambda e^{A(3)}_{\;\;\rho} \gamma^\sigma_{\;\;\nu} \left( \tg{T}^{\rho}_{\;\;\lambda\sigma}+\tg{T}^{\;\;\;\;\rho}_{\sigma\lambda}\right)\nonumber\\
    =&n_\lambda e^{A(3)}_{\;\;\rho} \gamma^\sigma_{\;\;\nu}\gamma^{\rho\epsilon} \left( 2\tg{K}^{\lambda}_{\;\;\epsilon\sigma}+\tg{T}^{\lambda}_{\;\;\epsilon\sigma}\right)\nonumber\\
    =&n_\lambda e^{A(3)}_{\;\;\rho} \gamma^\sigma_{\;\;\nu}\gamma^{\rho\epsilon} 2 \left(\tg{\Gamma}^{\lambda}_{\;\;\epsilon\sigma}-\accentset{\circ}{\Gamma}^{\lambda}_{\;\;\epsilon\sigma}\right)\nonumber\\
    =&e^{A(3)}_{\;\;\rho} \gamma^\sigma_{\;\;\nu}\gamma^{\rho\epsilon} 2 \left(\partial_\sigma n_\epsilon - \partial_\sigma n_\epsilon + \accentset{\circ}{\nabla}_\sigma n_\epsilon \right)\nonumber\\
    =&-2\accentset{\circ}{k}_\nu^{\;\;a}\,.
\end{align}
This confirms that the Lie derivative of the spatial metric and spatial tetrad for the Levi-Civita in GR and for the Weitzenb\"{o}ck gauge in TEGR are in fact directly related. 

Let us now derive the momentum and Hamiltonian constraints for TEGR. These are given here in their final form
\begin{align}
    \tg{T}^{(3)}_\rho \left(\tg{A}^{\rho}_{\;\;\chi}+\tg{A}_{\chi}^{\;\;\rho}\right)&-  \tg{A}^{\lambda\rho} \left(\tg{T}^{(3)}_{\lambda\chi\rho}+\tg{T}^{(3)}_{\rho\chi\lambda}\right)\nonumber\\
    &-2\tg{D}_{\chi} \tg{A} - \tg{D}_{\lambda}\tg{A}^{\lambda}_{\;\;\chi}=16\pi G S_\lambda\,,\label{eq:gr_momentum_constraint_TEGR}
\end{align}

\begin{align}
    2\tg{T}^{\lambda}_{(3)}\tg{T}_{\lambda}^{(3)}+&\frac{1}{2}\tg{A}^{\rho\sigma}\left(\tg{A}_{\rho\sigma}+\tg{A}_{\sigma\rho}\right) - \tg{A}^2 + 2 \tg{D}_\lambda \tg{T}^{\lambda}_{(3)}\nonumber\\
    &+ 4 \tg{D}_\rho \left(\frac{1}{\alpha}\partial^{\rho}_{(3)}(\alpha)\right)+\tg{T}^{(3)}=-16 \pi G \rho\,.\label{eq:gr_hamiltonian_constraint_TEGR}
\end{align} 
We finally consider the time vector $t^\lambda$ which can be written as \cite{baumgarte} 
\begin{equation}
    t^\lambda = \alpha n^\lambda + \beta^\lambda\,,
\end{equation}
where $\beta^\lambda$ is the shift vector. Taking the Lie derivative of $\tg{A}^A_{\;\;\nu}$ along the time vector gives 
\begin{equation}
    \mathcal{L}_{t}\tg{A}^A_{\;\;\nu}=\alpha \mathcal{L}_{n}\tg{A}^A_{\;\;\nu} +\mathcal{L}_{\beta}\tg{A}^A_{\;\;\nu}\,,\label{eq: t=an+b}
\end{equation}
where general Lie derivative properties have been used \cite{yano1957theory,baumgarte}. What is left is to consider the second term of Eq.~(\ref{eq: t=an+b}). Expanding this term, one obtains the following
\begin{align}
    \mathcal{L}_{\beta}\tg{A}^A_{\;\;\nu}=\beta^\lambda \partial_\lambda \left( \tg{A}^A_{\;\;\nu} \right) +  \tg{A}^A_{\;\;\lambda} \partial_\nu \left( \beta^\lambda \right)\,.
\end{align}
Similarly the spatial tetrad evolution equation becomes
\begin{align}
    \mathcal{L}_{t}e^{A(3)}_{\;\;\nu}&=\alpha \tg{A}^A_{\;\;\nu} + \beta^\lambda\left( \partial_\lambda \left\{ e^{A(3)}_{\;\;\nu} \right\} -   \partial_\nu \left\{e^{A(3)}_{\;\;\lambda} \right\}\right)\,.
\end{align}

\subsection{Gauss constraint and evolution equations in symmetric teleparallel gravity with the coincident gauge}\label{sec:stegr_decom}

We now consider STG within the coincident gauge which was described in Sec.~\ref{sec:STG_technical}. This theory is based on two field equations, one obtained by varying the Lagrangian with respect to the metric in Eq.~(\ref{eq:metric_FEs}), and another by varying it with respect to the connection in  Eq.~(\ref{eq:connection_FEs}). The coincident gauge for STEGR is important since it identically satisfies the connection field equations by setting the connection to be zero. This fact has a number of important implications. The most important of these is that the covariant derivative and the partial derivative become the same for this formalism.

The main Gauss equation for this construction is that for the metricity tensor. This was derived in Sec.~\ref{Sec:General_for} and it is given again here for ease of reference along with the Gauss constraint of  disformation tensor that is its direct consequence,
\begin{align}
    \gamma^{\lambda}_{\;\;\rho}\gamma^{\beta}_{\;\;\mu}\gamma^{\alpha}_{\;\;\nu}\stg{Q}_{\lambda\beta\alpha} = \stg{Q}_{\rho\mu\nu}^{\;(3)}\,,\\
    \gamma^{\lambda}_{\;\;\rho}\gamma^{\beta}_{\;\;\mu}\gamma^{\alpha}_{\;\;\nu}\stg{L}_{\lambda\beta\alpha} = \stg{L}_{\rho\mu\nu}^{(3)}\,.\label{eq:non-metricity_All_Gauss}
\end{align}
Let us  now move on to the evolution equations for STEGR. We start by introducing a modified version of the field equations in STEGR based on the fact that the evolution equations, in this case, are purely based on the disformation tensor, that is 
\begin{align}\label{eq: modified STGR FE}
    \partial_\alpha {\stg{L}^\alpha}_{\mu\nu}=&\;\Theta_{\mu\nu}-\frac{1}{2}g_{\mu\nu}\Theta+\stg{L}_{\sigma\;\;\alpha}^{\;\;\sigma}{\stg{L}^\alpha}_{\mu\nu}\nonumber\\
    &+\frac{1}{2}\partial_{(\mu|} {\stg{L}^\epsilon}_{\epsilon|\nu)}+{\stg{Q}^{\sigma\epsilon}}_\nu \stg{L}_{\sigma\epsilon\mu} - \frac{1}{2}  \stg{Q}_{\mu\nu\epsilon}{\stg{L}_\nu}^{\sigma\epsilon}\,.
\end{align}
Here, the scalar metricity is replaced by a combination of its definition and an expression obtained through the contraction of the metric filed equations
\begin{align}
    \stg{Q}=&\stg{L}^{\;\;\lambda}_{\lambda\;\;\alpha}\left({\stg{L}^{\alpha\epsilon}}_\epsilon-{\stg{L}_\epsilon}^{\epsilon\alpha}\right)+\partial_\alpha\left({\stg{L}_\epsilon}^{\epsilon\alpha}-{\stg{L}^{\alpha\epsilon}}_\epsilon\right)-\Theta\,.
\end{align}
The next step is to consider the evolution equations for STEGR within the coincident gauge. The generalized evolution equations Eqs.~(\ref{eq:firstlieofmetric_gen},\ref{eq:doublelieofmetric_nonmeticity}) thus become
\begin{align}
    \mathcal{L}_{n}\gamma_{\mu\nu}=&\stg{B}_{(\mu\nu)}\nonumber\\
    =&2\stg{B}_{\mu\nu}\,,\\
    \mathcal{L}_{n}\stg{B}_{\mu\nu}=&-\gamma^{\alpha}_{\;\;\nu}
    \gamma^{\beta}_{\;\;\mu}\partial_\lambda {\stg{L}^\lambda}_{\alpha\beta}+\partial^{(3)}_\lambda \stg{L}^{\lambda}_{\;\;\mu\nu}\nonumber\\
    &+\gamma^{\alpha}_{\;\;\nu}
    \gamma^{\beta}_{\;\;\mu} \stg{a}_\sigma {\stg{L}^\sigma}_{\alpha\beta}-\gamma^{\epsilon}_{\;\;\left(\nu\right|}
    \gamma^{\chi}_{\;\;\left|\mu\right)}\stg{B}_{\alpha\chi}n_\sigma{\stg{Q}_\epsilon}^{\sigma\alpha}\,.
\end{align}
After substituting the metric field equations into the second evolution equation, expressing all of the right hand side in spatial tensors, and writing the energy-momentum tensor and scalar in terms of the spatial stress $S_{\mu\nu}$, its trace $S$ and the density $\rho$, we obtain the final form of the second evolution equation in STEGR, that is 
\begin{align}
    \mathcal{L}_{n}\stg{B}_{\mu\nu}=&-8\pi G\left[S_{\mu\nu}-\frac{1}{2}\gamma_{\mu\nu}\left(S-\rho\right)\right]\nonumber\\
    &+\frac{1}{2}\stg{Q}_{\epsilon\;\;\sigma}^{\;\;\sigma (3)}\stg{L}^{\epsilon (3)}_{\;\mu\nu}+\frac{2}{\alpha}\partial^{(3)}_\epsilon \left(\alpha\right)\stg{L}^{\epsilon (3)}_{\;\mu\nu} +\partial^{(3)}_\sigma \stg{L}^{\sigma (3)}_{\;\mu\nu}\nonumber\\
   &-\stg{B}_{\mu\nu}\stg{B}-\frac{1}{2}\partial_{(\nu|}^{(3)}\stg{L}^{\sigma (3)}_{\;\sigma|\mu)}+\frac{1}{2}\partial_{(\nu|}^{(3)}\left(\frac{1}{\alpha}\partial_{|\mu)}^{(3)}\alpha\right)\nonumber\\
   &-\frac{1}{2}\stg{Q}^{\lambda\sigma(3)}_{\;\;\;\;\;\nu}\stg{Q}^{(3)}_{[\lambda\sigma]\mu}-\frac{1}{2}Q^{(3)}_{\mu\lambda\sigma}\stg{Q}^{(3)\lambda\sigma}_{\;\nu}+{\stg{B}^\sigma}_{(\nu|}\stg{B}_{\sigma|\mu)}\nonumber\\
   &-\frac{1}{\alpha^2}\partial^{(3)}_\nu\alpha \;\partial^{(3)}_\mu\alpha\,.
\end{align}
The Hamiltonian and momentum constraints for STEGR can then be derived through contracting Eq.~(\ref{eq: modified STGR FE}) in the standard way, and simplifying the results to obtain the respective final forms of the constraints 
\begin{align}
    16\pi G \rho =& -\frac{2}{\alpha^2}\partial^{(3)}_\nu(\alpha)\partial^\nu_{(3)}(\alpha) + \frac{1}{2}\stg{Q}^{\;\;\sigma (3)}_{\epsilon\;\;\sigma}\stg{L}^{\epsilon\;\;\nu}_{\;\nu(3)}+\gamma^{\mu\nu}\partial^{(3)}_\epsilon \stg{L}^{\epsilon(3)}_{\;\;\mu\nu}-\stg{B}^2-\partial^\nu_{(3)}\stg{L}^{\epsilon(3)}_{\;\;\epsilon\nu}\nonumber\\
    &\;-\frac{1}{2}\stg{Q}^{\lambda\sigma\nu}_{(3)}\stg{Q}^{(3)}_{\left[\lambda\sigma\right]\nu}-\frac{1}{2}\stg{Q}^{\lambda\sigma\nu}_{(3)}\stg{Q}^{(3)}_{\lambda\sigma\nu}+\stg{B}^{\sigma\nu}\stg{B}_{\sigma\nu}\,,\label{eq:stegr_con1}\\\nonumber\\
    8\pi G S_\beta=&\;\frac{1}{2}\stg{B}^{\epsilon\lambda}\stg{Q}^{(3)}_{\beta\epsilon\lambda} + \stg{L}^{\theta\rho(3)}_{\;\;\rho}\stg{B}_{\beta\theta}-\gamma^{\sigma\rho}\partial_\beta^{(3)}\left(\stg{B}_{\rho\sigma}\right)+\gamma^{\sigma\rho}\partial_\rho^{(3)}\left(\stg{B}_{\beta\sigma}\right)\,.\label{eq:stegr_con2}
\end{align}
Similar to the torsional case, we now consider the time vector and take the Lie derivative of $B_{\mu\nu}$ along the time vector, so that
\begin{equation}\label{eq: LtB = aLnB+LbB}
    \mathcal{L}_{t}\stg{B}_{\mu\nu}=\alpha \mathcal{L}_{n}\stg{B}_{\mu\nu} +\mathcal{L}_{\beta}\stg{B}_{\mu\nu}\,,
\end{equation}
where general Lie derivative properties have been used \cite{yano1957theory,baumgarte}. Expanding the second term of Eq.~(\ref{eq: LtB = aLnB+LbB}), one obtains the following
\begin{align}
    \mathcal{L}_{\beta}\stg{B}_{\mu\nu}=\beta^\lambda \partial^{(3)}_\lambda \left( \stg{B}_{\mu\nu} \right) +  \stg{B}_{\mu\lambda} \partial^{(3)}_\nu \left( \beta^\lambda \right)+\stg{B}_{\lambda\nu} \partial^{(3)}_\mu \left( \beta^\lambda \right)\,.\label{eq:stegr_evo1}
\end{align}
Similarly the spatial metric evolution equation becomes
\begin{align}
    \mathcal{L}_{t}\gamma_{\mu\nu}&=\alpha 2 \stg{B}_{\mu\nu} + \beta^\lambda\left( \partial^{(3)}_\lambda \left\{ \gamma_{\mu\nu} \right\} -   \partial^{(3)}_\nu \left\{\gamma_{\mu\lambda} \right\}-   \partial^{(3)}_\mu \left\{\gamma_{\lambda\nu} \right\}\right)\,.\label{eq:stegr_evo2}
\end{align}
Thus, this completes the transformation of the field equations into a set of four equations, two of which are evolution equations while the other two are constraint equations.

\section{Discussion and Conclusions}\label{sec:conclu}

In this paper, we have further probed the torsional and nonmetrical sides of GR through TEGR and STEGR, respectively. While  TG and STG rely on different constructions of gravity, as described in Sec.~\ref{sec:TG_STG_intro}, they both produce field equations that are dynamically equivalent to standard GR. For this reason it is crucial to study the entire trinity of ways in which GR can be written since different formalisms may offer better solutions for different issues of the theory.

Specifically, we presented a $3+1$ decomposition of both TEGR and STEGR in addition to the well-known GR decomposition. We first reconsider the $3+1$ decomposition strategy without assuming a connection by which a particular geometry can centre into the construction. We do this in Sec.~\ref{Sec:General_for} where we take a general decomposition of the metric and the tetrad and lay out the general strategy of this approach by describing the Lie derivatives and how they will lead to the evolution and constraints equations. 

Then, in Sec.~\ref{sec:specifics}, we describe our main results where we first present the well known $3+1$ decomposition within GR.  See Sec.~\ref{sec:gr_decom}. This analysis leads to two evolution equations and two constraint equations which are a set of equations that can fully determine any ensuing system analogous to the actual field equations. In Sec.~\ref{sec:tegr_decom}, we extend this strategy to the TEGR version of GR in which the fundamental dynamical variable becomes the tetrad (in the Weitzenb\"{o}ck gauge) which adds an additional layer of difficulty to the issue since this acts both on the general manifold and on the associated Minkowski space. In particular, we have  considered both holonomic and nonholonomic coordinates and transformations. In this part of the work,  we showed that this decomposition can take place and lay out the key elements of how this can be done for the evolution equations and highlighted the symmetry-asymmetry issue within them. We also present momentum and Hamiltonian constraint equations which complete the system. Altogether this gives a construction of TEGR in which numerical analyses, using this $3+1$ decomposition, can be carried out.

The STG version of GR is then probed in Sec.~\ref{sec:stegr_decom} where STEGR is fully decomposed using a $3+1$ approach. We do this in the coincident gauge so that the only dynamical variable is the metric tensor. Since we are again reliant on the metric tensor, the approach has some resemblance to the approach in standard GR, meaning that we can use some analogous relations in this setting. Indeed, we derive the evolution equations for STEGR in Eqs.~(\ref{eq:stegr_evo1},\ref{eq:stegr_evo2}), while the evolution equations are written explicitly in Eqs.~(\ref{eq:stegr_con1},\ref{eq:stegr_con2}). Thus, as in GR,  we can rewrite the field equations in a genuine $3+1$ decomposed structure which will be more amenable to numerical analyses.

The $3+1$ decompositions presented here can be further developed for implementation in numerical settings for specific scenarios. It may turn out that the novel approaches of TEGR and STEGR may  provide some advantage over the traditional method by which this formalism is implemented. One would expect this since both TEGR and STEGR remove boundary elements from their action whereas GR retains these elements. Since these three constructions of gravity form a trinity of the same underlying dynamical equations, they are guaranteed to be equivalent. These decomposition approaches are also  interesting because they may provide a way in which modified TG or STG gravity theories may be decomposed in a similar fashion in future works. Furthermore, the ADM couterparts for TG and STG can be extremely useful for achieving the canonical quantization of these theories also in view of Quantum Cosmology applications \cite{Book}.

\begin{acknowledgements}
JLS would like to acknowledge support from Cosmology@MALTA which is supported by the University of Malta. SC acknowledges the support of Istituto Nazionale di Fisica Nucleare (INFN), iniziative specifiche QGSKY and MOONLIGHT2. The authors also acknowledge also funding from ``The Malta Council for Science and Technology'' in project IPAS-2020-007.
\end{acknowledgements}

\bibliographystyle{utphys}

\providecommand{\href}[2]{#2}\begingroup\raggedright\endgroup

\end{document}